\documentclass[iop,revtex4]{emulateapj}
\usepackage{ifpdf}
\usepackage{hyperref}
\usepackage{amsmath}

\ifx \href \undefined \newcommand{\href}[2]{{\ttfamily #2}} \fi
\ifx \spc \undefined \newcommand{\spc}{\,} \fi
\ifx \doi \undefined
\newcommand{\doi}[1]{ DOI:\href{http://dx.doi.org/\detokenize{#1}}{\ttfamily
  \detokenize{#1}}\spc}
\fi
\ifx \eprint \undefined
\newcommand{\eprint}[2]{
  {#1}:\href{http://\detokenize{#1}.org/abs/\detokenize{#2}}{\ttfamily
  \detokenize{#2}}\spc}
\fi
\ifx \arxiv \undefined
\newcommand{\arxiv}[1]{ \eprint{arXiv}{#1}}
\fi
\ifx \vixra \undefined
\newcommand{\vixra}[1]{ \eprint{viXra}{#1}}
\fi
\ifx \ads \undefined
\newcommand{\ads}[1]{
  ADS:\href{http://adsabs.harvard.edu/abs/\detokenize{#1}}{\ttfamily
  \detokenize{#1}}\spc}
\fi

\newcommand{\tentothe}[1]{\ifmmode {10^{#1}} \else {$10^{#1}$} \fi}
\newcommand{\tten}[1]{\ifmmode {\times 10^{#1}} \else {$\times 10^{#1}$} \fi}
\newcommand{\unit}[1]{\ifmmode {\rm\ #1} \else {$\rm #1$} \fi}
\newcommand{\AU}{\unit{A.U.}}

\newcommand{\Kepler}{{\em Kepler}}
\newcommand{\WISE}{{\em WISE}}

\newcommand{\IRAS}{{\em IRAS}}

\newcommand{\K}{\unit{K}}

\newcommand{\degrees}{\ifmmode{^{\circ}}\else{$^{\circ}$}\fi}
\newcommand{\degree}{\ifmmode{^{\circ}}\else{$^{\circ}$}\fi}

\newcommand{\doublet}{\ifmmode {\lambda\lambda} \else {$\lambda\lambda$} \fi}

\newcommand{\kg}{\unit{kg}}

\newcommand{\lt}{\unit{<}}

\newcommand{\milligram}{\unit{mg}}

\newcommand{\persec}{\unit{s^{-1}}}
\newcommand{\ergpersec}{\unit{erg\ \persec}}
\newcommand{\persqrcm}{\unit{cm^{-2}}}
\newcommand{\persquarekm}{\unit{km^{-2}}}

\newcommand{\quarter}{\ifmmode {\frac{1}{4}} \else {$\frac{1}{4}$} \fi}

\newcommand{\sciam}{Scientific American}
\newcommand{\science}{Science}
\newcommand{\singlet}{\ifmmode {\lambda} \else {$\lambda$} \fi}

\newcommand{\squarekm}{\unit{km^2}}

\newcommand{\ifms}[1]{}	
\newcommand{\ifpp}[1]{#1}	
\newcommand{\ifbw}[1]{}		
\newcommand{\ifcol}[1]{#1}	

\slugcomment{Submitted to {\it The Astrophysical Journal}}

\shorttitle{Technology in Light Curves}
\shortauthors{Korpela, Sallmen \& Greene}

\begin{document}


\title{Modeling Indications of Technology in Planetary Transit Light Curves -- Dark-side illumination.}


\author{Eric J. Korpela}
\affil{Space Sciences Laboratory, University of California at Berkeley, Berkeley, CA, 94720}
\email{korpela@ssl.berkeley.edu}
\author{Shauna M. Sallmen, Diana Leystra Greene}
\affil{University of Wisconsin - La Crosse, La Crosse, WI 54601}

\begin{abstract}
We analyze potential effects of an 
extraterrestrial civilization's use of orbiting mirrors to
illuminate the dark side of a synchronously rotating planet
on planetary transit light curves.  Previous efforts to
detect civilizations based on side effects of planetary-scale engineering
have focused on structures affecting the host star output (e.g. Dyson
spheres).  However, younger civilizations
are likely to be less advanced in their engineering efforts, yet still
capable of sending small spacecraft into orbit.  Since M dwarfs are the
most common type of star in the solar neighborhood, 
it seems plausible
that many of the nearest habitable planets orbit dim, low-mass M stars, 
and will be in synchronous rotation. Logically,
a civilization evolving on such a planet may be inspired to illuminate
their planet's dark side by placing a single large mirror at the L2
Lagrangian point, or launching a fleet of small thin mirrors into
planetary orbit. 
We briefly examine the requirements and engineering
challenges of such a collection of orbiting mirrors, then explore their 
impact on transit light curves.  We incorporate
stellar limb darkening and model a simplistic mirror fleet's effects 
for transits of Earth-like ($R$ = 0.5 to 2 $R_{\rm Earth}$) planets which 
would be synchronously rotating for orbits within the habitable zone of their host star.   
Although such an installation is undetectable in \Kepler\ data, {\it JWST}
will provide the sensitivity necessary to detect a fleet of mirrors orbiting
Earth-like habitable planets around nearby stars.

\end{abstract}


\section{Introduction}

The detection by \Kepler\ \citep{marcy14,petigura2013} of multiple Earth-like planets in 
or near the habitable zones of late type stars has renewed interest in the
search for extraterrestrial intelligence (SETI).  Most 
prior searches concentrated on
detecting electromagnetic communications in radio/microwaves and optical/IR
bands \citep{vonkorff13,korpela11,howard07,tarter01,bowyer97,sagan75}. 
Such searches have operated intermittently since the 1960s and
continue today with greatly enhanced frequency coverage and detection
sensitivity.

\citet{freitas83} suggested that the first evidence we
find of an extraterrestrial civilization may be the detection of a
physical artifact rather than an electromagnetic signal.  In this paper
we investigate the potential for detection of a large fleet of small spacecraft
in orbit around a terrestrial exoplanet, for example, mirrors illuminating the
dark side of a planet in synchronous rotation.
Such a fleet could be observable through its effect
on the transit light curve. 

The remainder of the introduction motivates this scenario and
places it in context with conjectures regarding other extraterrestrial
engineering projects and our current understanding of exoplanet 
habitability.
In Section \ref{sect:methods} we explore plausible methods of dark-side 
illumination, discussing associated technical and engineering challenges 
in Section
\ref{sect:engineering}. Section \ref{sect:models} describes a simple 
model of
the effects of an orbiting mirror fleet on transit light curves. 
We present the resulting light curves 
in Section \ref{sect:sims}, focusing on the signature 
and magnitude of the mirrors' influence, and examining the effects 
of altering various stellar, planetary, and mirror fleet parameters.
We consider prospects for detection in Section \ref{sect:detect}.

\subsection{Motivation}

The most common type of star in our Galaxy is the M
dwarf. 
These stars have very long lives. Because of their low
luminosity, the habitable zone (HZ) around M dwarfs is narrow and
close to the star.  Due to tidal considerations, 
such planets 
will probably always present one face to the star, except in unusual cases. 
\citep{Tarter07}
Although young ($\lt 100$ Myr) dM stars have active
chromospheres and high UV output that could be damaging to biological organisms,
such activity decreases rapidly with
age \citep{engle09}. There is also a wide variation in activity 
at all ages \citep{shkolnik2014}, and its precise impact on exoplanet 
atmospheres, water content, magnetospheres and habitability is a 
subject of active research 
\citep[e.g.][]{mulders2014,tian2014,zuluaga2013,grenfell2012}. 
Since the specific conditions required for the evolution of an 
ecosystem are not well-defined, and given that M and K dwarfs are 
80\% of stars in the solar neighborhood \citep{henry06}, 
it seems plausible that most nearby 
habitable planets are in synchronous rotation in orbits around low-mass stars.
A significant fraction of M dwarfs do host small planets
within their habitable zones, with the detailed estimates depending
on the choice of HZ \citep{dc2013,gaidos2013,kopparapu2013}.

Life around an M dwarf could pose unique challenges for life,
and for the evolution of intelligence. But once a civilization has
either evolved {\em in planeta} or migrated from elsewhere, it
might wish to utilize portions of the planet that do not
receive illumination from the star.  After careful consideration of the climatic
effects, the first deliberate planetary scale
engineering project might be illumination of the planet's dark side. The most
straightforward method is via orbiting thin mirrors,
either placing a single large mirror at the L2 Lagrange point or, more likely, 
launching a
fleet of smaller mirrors into lower orbits. To provide significant illumination,
these mirrors must have a total reflective surface area 
approximately equal to the disk area of the planet. 
This could render them detectable in a planetary transit.

If the primary motivation is dark-side illumination, we anticipate such 
structures would be found primarily
around synchronously rotating planets, most likely orbiting
M stars.  As we discuss in Section \ref{subsect:HZ}, habitable worlds in 
certain orbits around earlier-type stars (e.g. G stars) might also be 
in synchronous rotation at the current time.
In addition, an alien civilization
might wish to utilize a previously uninhabited planet in their star
system, or simply ensure illumination of the night side of a planet
in non-synchronous rotation.  Other possible uses of fleets of comparable area
include photoelectric power generation.  Thus, although in what follows we focus our
attention on worlds with a perpetual dark side, 
such engineering projects might occur in other situations, and around 
many types of stars.

\subsection{Prior proposed megastructures}

\citet{dyson60} suggested that an extraterrestrial civilization might build 
a giant structure to capture and utilize most 
or all of the visible emissions of a star, radiating only waste heat.
A Dyson sphere 
would be detectable as a stellar-luminosity $\sim 300 \K$ featureless 
blackbody source if the structure completely encases the star, or as a 
star with a 
significant infrared (IR) excess in the case of a partial sphere.  If civilizations
capable of such feats of engineering were common, we could likely
detect them in IR surveys.  Searches of the IR sky sensitive to both complete 
\citep{carrigan09} and partial (\nocite{JugakuNish04}Jugaku \& Nishimura 2004; 
Conroy \& Werthimer
2003, unpublished\footnote{From
\href{http://seti.berkeley.edu/IR_Excess_Search}{http://seti.berkeley.edu/IR\_Excess\_Search} retrieved 09
July, 2015}; Globus et al. 2003, unpublished\footnote{Mentioned in
\citet{carrigan09}}  ) 
Dyson spheres have been conducted using the \IRAS\ dataset.  
While these have detected a few interesting candidates, none have
been shown to indicate intelligent origin.
Two high-sensitivity searches for Dyson spheres
are ongoing, one searching for IR signatures of Dyson
spheres in the \WISE\ survey \citep{wright14a, wright14b, griffith15},
and another examining \Kepler\
light curves for unusual transit eclipses that could
indicate a partial Dyson sphere  \citep{Marcy13_dyson}. 

A civilization advanced enough to trap most of the available output of
their host star ($\sim 4\tten{33} \ergpersec$) meets the definition
of Kardashev Type II on \citet{kardashev64}'s scale of technological 
advancement of a civilization.  A Type I civilization 
uses energy on the scale of current terrestrial civilization 
($\sim 4\tten{19} \ergpersec$).  Given the possibility of extinction,
it seems likely that Kardashev Type II civilizations
are much rarer than those capable of sending small spacecraft into
orbit, and that a continuum of possible engineering stages should exist.
We might consider a civilization with significant ability to
modify its planetary environment, but not capable of building \AU~scale
structures as Type I.5.

\citet{carrigan12} recently summarized several other potential 
archaeological signatures for civilizations at various levels on the 
Kardashev scale, concluding that apart from electromagnetic communications, 
these are generally undetectable
with current human technologies.  \citet{arnold05} suggests that a  
extraterrestrial civilization could deliberately create
planetary scale structures with unique shapes or periodicities that could 
be detected when the structure transits the star, but could 
be distinguished from a planetary transit with current technology.  Our 
proposed engineering project is of similar scale, but 
with a useful function besides communication and with a form that follows 
from its function. 

\section{Dark-side illumination utilizing mirrors}\label{sect:darkside}

\subsection{Methods of dark-side illumination}\label{sect:methods}

\begin{figure}[tbp]

\ifms{
\epsscale{0.6}
\ifpdf
\plotone{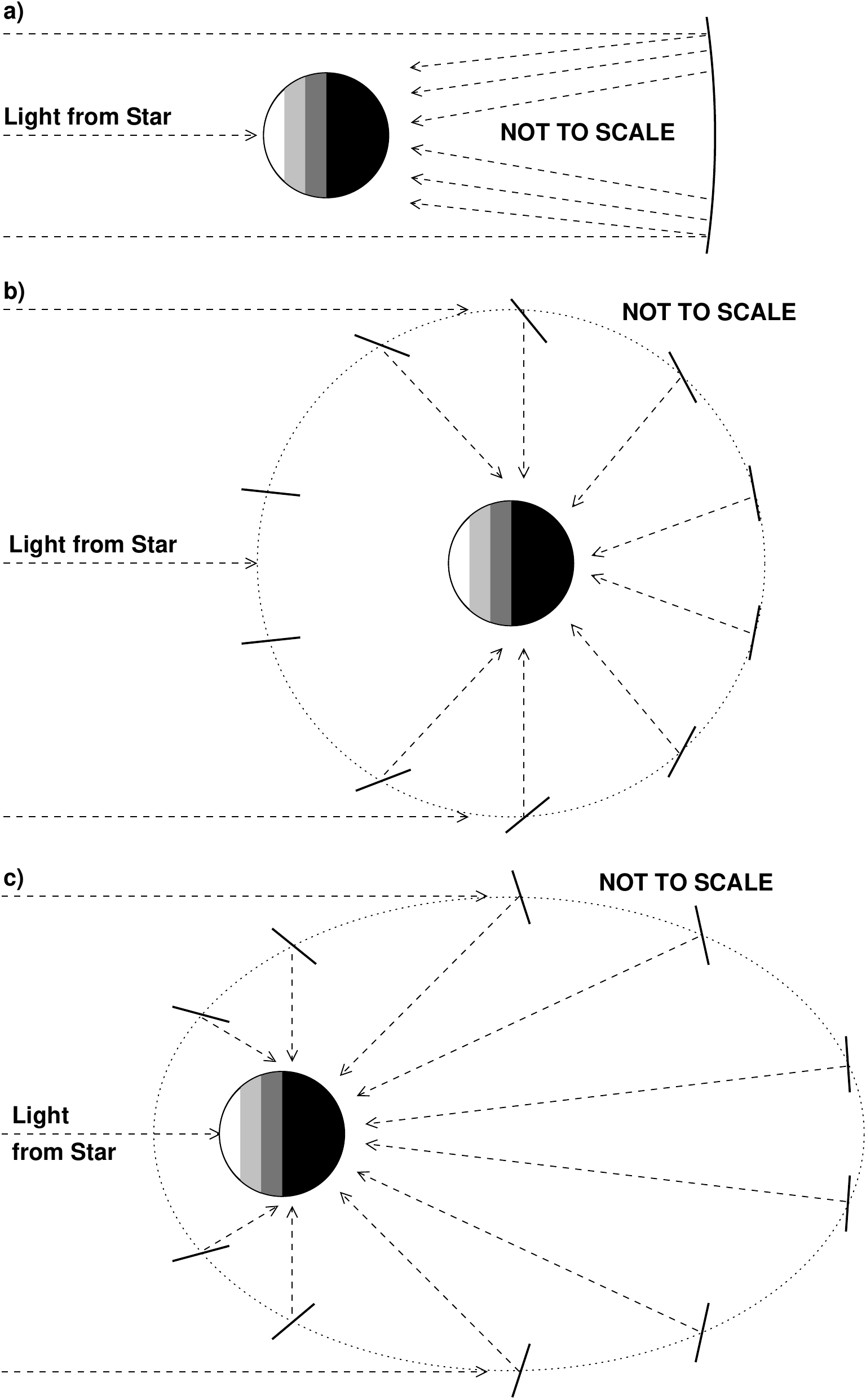}
\else
\plotone{fig1.eps}
\fi
}
\ifpp{
\ifpdf
\begin{center}
\includegraphics[width=0.9\columnwidth]{fig1.pdf}
\end{center}
\else
\begin{center}
\includegraphics[width=0.9\columnwidth]{fig1.eps}
\end{center}
\fi
}
\caption{Schematic illustration of three methods of dark-side illumination 
(not to scale). Planetary gray scale bands indicate different levels 
in stellar illumination.  In the three cross-sectional drawings, 
(a) shows a large circular or annular mirror stationed at the L2 Lagrange point,
(b) shows multiple small mirrors in circular orbits, 
(c) shows multiple small mirrors in elliptical orbits designed to maximize 
the duty cycle of the mirrors. 
\label{fig:methods}}
\end{figure}

There are multiple methods by which an alien civilization
could illuminate the dark side of a synchronously rotating planet.  
Conceptually,
the simplest is placing a single large mirror at the L2 Lagrange
point (Figure \ref{fig:methods}a).  To provide illumination equivalent
to the bright-side stellar illumination, it would need to be
a large disk or annulus somewhat larger than the planetary diameter.
Since the mirror is locked in the anti-stellar direction it would provide
non-uniform directional illumination on the ground, with regions near the
terminator receiving illumination at higher incidence angles than regions
near the anti-stellar point.  One notable difficulty 
is maintaining positional and directional
stability at the unstable L2 point.  A structure and control system
that could allow pointing and positional control of a 6000 km to 25000
km diameter mirror would be a significant engineering challenge.

In the case of a single large ($R_m > R_p$) mirror at the L2 Lagrange
point, the optical transit light curve will last longer and be deeper
than if only the planet blocked the starlight.  If the mirror has
been engineered so all intercepted light is redirected to the planet, and
none towards our line-of-sight, the planetary eclipse phase will be the
same as if the mirror were not present.  Thus the difference in transit
and eclipse time-span would be the main clue to the presence of
such a structure, but a similar timing effect could occur for a planet 
in a highly elliptical orbit with periastron at the time of eclipse 
and apastron at the time of
transit. 

A less technically challenging method would be to deploy a fleet
of independently steerable smaller mirrors. Figure
\ref{fig:methods}b schematically illustrates this possibility in cross-section for
mirrors in a circular orbit. Elongated mirror orbits, as depicted in
Figure \ref{fig:methods}c, might be preferable, since they spend
a greater fraction of their orbital period in a position to usefully
illuminate the planet's dark side.

For a fleet of many individual reflectors, the translucent fleet of
mirrors would block some starlight, deepening and lengthening the transit. 
Before modeling in detail the effects of a fleet of space-based mirrors on 
transit
light curves, we first discuss technical and engineering considerations
involved in its implementation.

\subsection{Technical \& Engineering Considerations}\label{sect:engineering}

Current materials technologies allow construction of metalized plastic
films with surface mass densities of $0.7-1.5 \milligram\persqrcm$;
neglecting supporting structures, the mass of a 1 \squarekm mirror would
be 7000 to 15000 kg \citep{bryant14}. Future
materials development (carbon nanotube fabrics: \citet{lima11}, space-based 
fabrication of metal films: \citet[][p.109]{wright92}) could 
result in materials 
with surface mass densities from 0.01 \milligram\persqrcm to as little as 
0.002 \milligram\persqrcm. This 
provides the potential to allow in situ creation 
of 1 \squarekm reflective surfaces with masses of 20 to 100 kg.
For these, structure and propulsion/control systems 
will likely be the dominant mass.  Each reflector could be
pointed based upon orbital position, and controlled independently and
autonomously.

We will not attempt a detailed discussion of the dynamics of such an
installation in this paper. However, we can utilize the planet's L2 
Lagrangian point to estimate the largest stable
orbits in the effective planetary gravitational well.
For a planet
in orbit at the inner edge of the HZ of an M8 star (as estimated below in
Section \ref{subsect:HZ}), the Lagrange point is beyond 10 planetary radii,
and is even more distant for earlier stellar types.
Therefore, for mirror flotillas
with orbits of modest extent ($\lesssim$ 5 planetary radii), the
mirror orbits should be relatively stable and controllable.

Providing a reflective area equivalent to Earth's disk area (1.3\tten{8}
\squarekm) requires deployment of more than 130 million
mirrors of 1 \squarekm each, or 1.3 million mirrors of 100 \squarekm.
At 1000 \kg \persquarekm, even a 100 \squarekm reflector lies within
current human launch capabilities.  However, even with
optimistic (\$2200/kg) estimates\footnote{In 2015, 
launch costs to geosynchronous transfer orbit on Falcon Heavy in 2015 
were estimated at \$14100/kg: 
\href{http://www.spacex.com/about/capabilities}{http://www.spacex.com/about/capabilities}, retrieved on May 16, 2015},
launch costs for such a fleet exceed $\$2.8\tten{14}$ USD$_{2015}$.

There will undoubtedly be significant environmental impacts due to
pollution from manufacturing, launch exhaust products and altering
the energy balance of the planet.  However, we leave
the details of engineering a climatic balance, combating pollution,
and the possible sacrifice of a decade\footnote{Earlier revisions of this
manuscript indicated 100 times higher cost, due to an aritmetic error}
of economic output to the
inhabitants of these worlds.

It is worth highlighting certain considerations affecting
implementation of dark-side illumination.
For example, heating the dark side would lower the day-night temperature
differential and slow day-to-night heat transfer, warming the day side.
The mirrors would also obscure some starlight for
day-side inhabitants. Optimally the mirrors  
could be turned edge-on to our transit line-of-sight
 quickly as they exit the zone of usefulness, in order to minimize
obscuration, unnecessary exposure to solar radiation, and forces that would
alter the orbit. However, some obscuration may be desirable to
reduce the stellar flux to the day side and keep planetary temperatures
from rising excessively.  Obscuration of the day side
isn't overly large even if the mirrors are not directed edge on, lying  
between $\sim R_p^3/R_m^3$ and $\sim R_p^2/R_m^2$ 
depending on the orbital arrangement, where $R_m$ is the orbital radius of the
mirror fleet. Satellites in a 3$R_p$ orbit would decrease
the flux by approximately 4 to 11\%, while a 10$R_p$ orbit reduces
flux by 0.1\% to 1\%.  

As a result of these and other considerations, the detailed distribution
and orientation of mirrors within the orbiting fleet are likely 
quite complicated.  The orbital distribution and variable incidence angle
of the mirrors would affect the actual mirror surface area required, as
well as the observed transmittance of the mirror fleet.  However, for most of
this paper we restrict ourselves to the simplest case and consider the
observable impact of a constant-transmittance fleet of mirrors surrounding an
extrasolar planet. Near the end of Section \ref{sect:sims}, we briefly
consider alternative transmittance models, while deferring more detailed
modeling of mirror fleet transmittance to another time.

\section{Modeling Effects of a Cloud of Mirrors}\label{sect:models}

An orbiting mirror fleet partially blocks starlight 
just before, during, and after planetary transit. 
We initially assume the fleet appears in cross-section as a
constant-transparency ring around the planet. Although 
unlikely to be completely accurate, these models highlight the first-order
effects of orbiting mirrors on transit light curves.
We also neglect the increased light reflected from the now-illuminated 
planetary dark side, an assumption validated in Section \ref{sect:2ndorder}.

We modeled fleets of orbiting mirrors in two ways. For our standard runs, 
we adapted the {\sc
exofast\_occultquad} routines \citep{eastman13} based on the algorithms
of \citet{MA02}. 
This package calculates light curves as a function of the normalized
separation of the star and planet centers ($z = d/R_{\rm star}$), which
for circular orbits is related to time ($t$; 0 at transit center) by 
$$z = \frac{a_P}{R_{\rm star}} \sqrt{(\sin\omega t)^2 + (\cos i \cos \omega
t)^2}$$ 
where $i$ is the orbit inclination and $a_P$ and $\omega$ describe the
radius and angular frequency of the planetary orbit. 

The {\sc exofast} routine is used to produce a light curve for a
transit by a planet without mirrors ($LC_P$), and again to produce a
light curve for a transiting object the size of the mirror fleet ($LC_O$).
The fraction of starlight blocked in each case is $B_P = 1 - LC_P$ and 
$B_O = 1 - LC_O$, respectively. The light blocked by
the transparent mirror ring (absorptance = $\alpha$ = 1 - transmittance) 
is therefore $B_M = \alpha (B_O - B_P)$ and the light curve for a 
planet encircled by the mirror fleet is 
\begin{equation} LC_{\rm tot} = 1 - B_M - B_P = LC_P (1 - \alpha) +
\alpha LC_O \label{eq:mirr_trans} 
\end{equation}
The absorptance $\alpha$ and optical depth $\tau$ are 
nearly equivalent for $\alpha \lesssim 0.1$; however this is not
necessarily valid for our simulations.


We assume the effective
cross-sectional area of the constant-transparency mirror ring ($\alpha A_m$)
equals the cross-sectional area of the planet ($A_P$), resulting in an absorptance 
\begin{equation} 
\alpha = \frac{A_P}{A_m} = \frac{\pi R_P^2}{\pi(R_m^2 - R_P^2)}= \frac{1}{(\frac{R_m}{R_P})^2 - 1}.\label{eq:alpha}\end{equation}
This is reasonable because any mirrors between Earth and the planet
are not illuminated, so do not contribute to the area needed to
illuminate the dark side at day-side levels. We also assume 
mirrors are turned edge-on quickly as they exit the zone of usefulness.
i.e. Any mirrors in a position
that renders them incapable of illuminating the planet's dark side
are edge-on so don't contribute to the fleet's absorptance.

Although our standard simulations
assume the mirror fleet provides constant absorptance, we 
wished to explore the importance of this simplification. 
We therefore created
software to generate a two-dimensional array representing the absorptance of 
the planet and mirror fleet, and another representing the star. For all
runs, the planet radius was set to 100 pixels. We then shifted the 
relative positions of the two arrays, calculated the transmitted light 
for each location, and interpolated the resulting light curve onto the same
temporal grid used with {\sc EXOFAST}.  This brute-force modeling approach
is significantly slower than that already described, but in all constant-absorptance test 
cases the light curves generated by the two methods differed by at most 
0.05\% of the transit depth, within the range expected due to 
pixelization errors. We used this approach to simulate transits 
for configurations with non-constant absorptance profiles (see Section \ref{sect:mirreff}).

\subsection{Parameterization}\label{subsect:HZ}

Table \ref{table:MS_stars} summarizes the properties of stars used in our
transit models. Columns 1-5 contain the spectral type and 
basic stellar properties from \citet{zombeck}.  
Values for surface gravity log($g$) were estimated two ways.
For Method A we retrieved
information on \Kepler\ stars hosting exoplanets from 
the Exoplanet Data Explorer \citep{wright2011} at 
\href{http://exoplanets.org/table}{http://exoplanets.org/table},
fit the data
with $log(g) = c_1 + c_2 T_{\rm eff}$, then utilized this
function to estimate $log(g)$. For Method B we scaled the 
surface gravity from the solar value, assuming $g \propto M/R^2$. 
In several test runs, the differences in the resulting light curves 
were at most 1\% of the transit depth, and 
generally less than 0.2\%.  For all of our standard runs we used the values 
from Method A, as shown in column 6.

\begin{table*}[tb]
\begin{center}
\caption{Stellar Parameters\label{table:MS_stars}}
\begin{tabular}{cccccc}
\hline
Sp Type & T$_{\rm eff} (K)$ & M/M$_{\rm sun}$ & R/R$_{\rm sun}$ & L/L$_{\rm sun}$
        & log($g$) \\
\hline\hline
M8 & 2660 & 0.10 & 0.13 & 0.0008 & 4.96 \\
M5 & 3120 & 0.21 & 0.32 & 0.0079 & 4.87 \\
M0 & 3920 & 0.47 & 0.63 & 0.063 & 4.72 \\
K5 & 4410 & 0.69 & 0.74 & 0.16 & 4.62 \\
K0 & 5240 & 0.78 & 0.85 & 0.40 & 4.46 \\
G5 & 5610 & 0.93 & 0.93 & 0.79 & 4.39 \\
G2 & 5780 & 1.00 & 1.00 & 1.00 & 4.35 \\
G0 & 5920 & 1.10 & 1.05 & 1.26 & 4.33 \\
\hline
\end{tabular}
\end{center}
\end{table*}

For our standard runs, we assumed heavy-element abundances half that 
of the Sun (log([Fe/H]$_{\rm ratio}$) = $-$0.3), and calculated
quadratic limb-darkening parameters $\gamma_1$ and $\gamma_2$ using the 
interface of \citet{eastman13} (
\href{http://astroutils.astronomy.ohio-state.edu/exofast/limbdark.shtml}{http://astroutils.astronomy.ohio-state.edu/exofast/limbdark.shtml}),
which interpolates over the values of \citet{ClaretBloemen11}.
The relatively low element abundance value was chosen because we anticipate
advanced civilizations around older stars which tend to be metal poor.

We assumed circular planetary orbits in all our models.  
Edge-on orbits were standard, 
but we also explored other orbit inclinations.
To determine appropriate orbit sizes, we estimated each star's habitable zone
using the 
{\it optimistic} values of \citet{kopparapuetal13}, accessed via
\href{http://depts.washington.edu/naivpl/content/hz-calculator}
{http://depts.washington.edu/naivpl/content/hz-calculator}. 
We note that substantial disagreement exists on the precise range of
orbits which lie in a star's HZ. For a detailed discussion,
see \citet{petigura2013}, who compare the conservative (rather than
optimistic) choices made by \citet{kopparapuetal13}
with those of other authors \citep{kasting93,zsom2013,PandG2011}.  However,
in this paper we are concerned mainly with illustrating the potential
impacts of a fleet of orbiting mirrors on transit light curves,
using one plausible choice for the HZ.

Table \ref{table:planets} summarizes the properties of planets 
used in our transit models. For each stellar spectral type 
(column 1), the inner edge (HZ$_{\rm in}$) and center 
(HZ$_{\rm mid}$) of the chosen HZ are indicated 
in columns 2 and 5. Corresponding planetary orbit periods for circular orbits
(P$_{\rm orb,in}$ and P$_{\rm orb,mid}$), determined from stellar 
mass and orbit size, are in columns 3 and 6.  Because we anticipate 
mirror fleets might be implemented around planets in synchronous rotation, 
columns 4 and 7 contain estimates of the tidal synchronization timescale 
($\tau_{\rm syn}$) for Earth-like ($M=M_E$, $R = R_E$) planets in these orbits.
Planets which would not synchronize within $\sim 10$ Gyr are marked with an asterisk ($^*$).

\begin{table*}[tb]
\begin{center}
\caption{Planet Parameters\label{table:planets}}
\begin{tabular}{cccccccc}
\hline
Sp Type & HZ$_{\rm in}$ (AU) & P$_{\rm orb,in}$ (d) & $\tau_{\rm syn,in}$ (Gyr) 
    & HZ$_{\rm mid}$ (AU) & P$_{\rm orb,mid}$ (d) & $\tau_{\rm syn,mid}$ (Gyr)\\
\hline\hline
M8 & 0.023 & 3.27 & 3 $\times 10^{-7}$ & 0.043 & 9.24 & $1 \times 10^{-5}$ \\
M5 & 0.073 & 14.8 & 7 $\times 10^{-5}$ & 0.13 & 37.4 & 2 $\times 10^{-3}$ \\
M0 & 0.20 & 47.7 & 6 $\times 10^{-3}$ & 0.36 & 115 & 0.2 \\
K5 & 0.32 & 79.6 & 5 $\times 10^{-2}$ & 0.55 & 179 & 1 \\
K0 & 0.48 & 138  & 0.4 & 0.83 & 313 & 11\tablenotemark{*} \\
G5 & 0.67 & 208  & 2 & 1.13 & 455 & 50\tablenotemark{*} \\
G2 & 0.75 & 237  & 4 & 1.26 & 517 & 90\tablenotemark{*} \\
G0 & 0.84 & 268  & 6 & 1.40 & 577 & 100\tablenotemark{*} \\
\hline
\end{tabular}
\tablecomments{$^*$ Earth-like planets in these orbits do not synchronize 
their rotation within 10 Gyr.}
\end{center}
\end{table*}

To estimate $\tau_{\rm syn}$ we used Equation 3 
of \citet{ForgetLeconte2013}, based on the equilibrium theory
of tides \citep{Darwin1880,Hut1981,Leconte2010}. 
We followed their lead in 
estimating tidal dissipation values according to
Lunar Laser Ranging experiments of \citet{nds_las1997}, which yields
\begin{equation}
\tau_{\rm syn}({\rm Gyr}) = 21.71 \frac{(M_p/M_E) a({\rm AU})^6}{G (M_s/M_{\rm sun})^2(R_p/R_E)^3} 
\end{equation}
Terrestrial planets at the {\it inner edge} of the chosen habitable zone 
(HZ$_{\rm in}$)
synchronize their rotation within 6 Gyr for all G, K and M 
main sequence stars. However, $\tau_{\rm syn}$ is longer for terrestrial planets elsewhere 
in the HZ, For example, for planets at its {\it center} 
(HZ$_{\rm mid}$), synchronous rotation is achieved within $\sim 1$ Gyr for main
sequence stars K5 and later, but not within 10 Gyr
for stars K0 and earlier.  Tidal synchronization time-scales are uncertain
at best. Estimates made by \citet{kasting93} suggest that for
planets orbiting at HZ$_{\rm mid}$, only those around
M stars would be in synchronous rotation at the current epoch, while those
orbiting at HZ$_{\rm in}$ around most K stars
could also be in this state.  Although detailed estimates
vary, planets in an M star's HZ
are extremely likely to be in synchronous rotation after a several Gyr biological evolution 
timescale.  However note that a planet's climate and habitability 
may also be altered by effects such as tidal heating or orbital migration 
\citep[see][]{barnes08,barnes13}.
As they are most likely to be in synchronous rotation, we focus on planets
around M stars for the majority of our models, but do explore other options, since
Earth-like planets orbiting a G star near HZ$_{\rm in}$ might be in synchronous
rotation after sufficient time,
and as noted earlier, 
civilizations on Earth analogues might choose to utilize a fleet of mirrors for reasons
other than dark-side illumination of a synchronously rotating planet.

In each orbit, we simulated transits for planetary radii of $R_P/R_{\rm Earth}$ = [0.5, 1, 2], 
corresponding to $M/M_E$ = [0.125,1,8] if all have the same average density
as Earth, or $M/M_E$ = [0.53,1, 5.1] using the relationships of 
\citet{weiss_marcy14}.
Effects on the transit due to brightness of the planet's dark side were ignored.
For each planet, we simulated mirror fleets extending to three different 
distances from the planet center: $R_m / R_P$ = [2, 3, 10]. Mirror
fleets that are substantially smaller ($R_m/R_P = \sqrt{2}$) will be 
nearly opaque according to equation \ref{eq:alpha}, while those that
are substantially larger may have less stable orbits and may suffer from
inefficiencies in redirecting starlight.

For all runs, in addition to determining the light curve for the planet
surrounded by its mirror fleet ($P+M$), we also determined the light
curve for the planet alone ($P$) and for a larger planet ($LP$) sans mirrors
that produces the same eclipse depth as $P+M$.

\section{Results of Simulations}\label{sect:sims}

Overall, mirror-induced perturbations to the transit shape are
similar for all situations.  We begin by presenting one model as a reference,
then explore the effects of varying stellar, planetary, and
mirror fleet parameters.  We chose a large terrestrial planet orbiting
an M5 star as our reference, because these stars and planetary sizes are
relatively common, and a relatively large planet will create a deeper
transit with a longer entrance time, making it easier to analyze the
transition region potentially affected by the mirror fleet.

\begin{figure}[tbp]
\ifpdf
\plotone{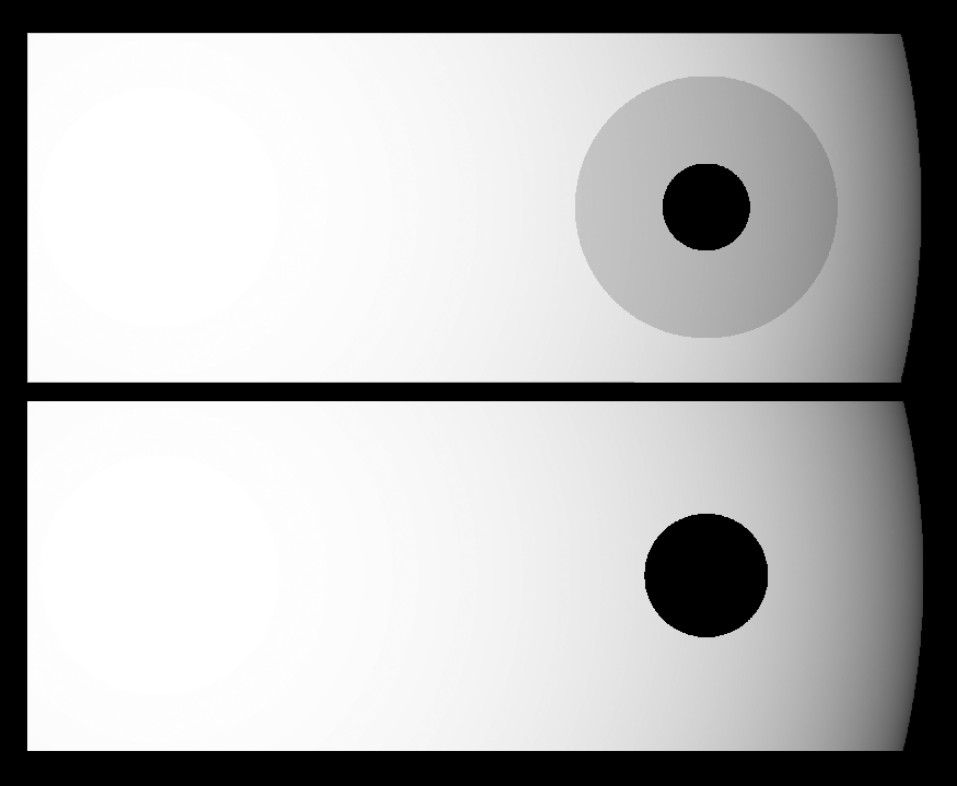}
\else
\plotone{fig2.eps}
\fi
\caption{{\it Top panel}: Simulated image of a planet with $R_P = 2 R_{\rm Earth}$,
surrounded by a constant-absorptance mirror fleet with $R_m = 3R_P$, transiting
in front of an M5 star.  Only a strip of the right half of the star is shown. 
{\it Bottom panel}: Same as above, but for an isolated planet large enough to produce
a transit of the same depth.
\label{fig:M5_img}}
\end{figure}

\begin{figure}[tbp]
\ifbw{
\ifpdf
\includegraphics[width=0.99\columnwidth,keepaspectratio=true]{fig3_BW.pdf}
\else
\includegraphics[width=0.99\columnwidth,keepaspectratio=true]{fig3_BW.eps}
\fi
}
\ifcol{
\ifpdf
\includegraphics[width=0.99\columnwidth,keepaspectratio=true]{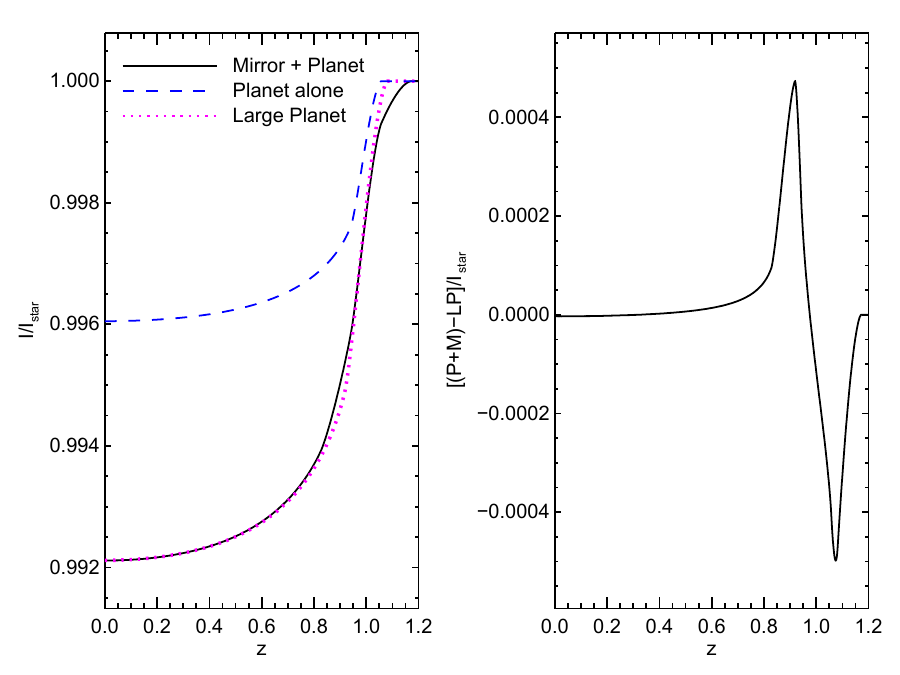}
\else
\includegraphics[width=0.99\columnwidth,keepaspectratio=true]{fig3_col.eps}
\fi
}
\caption{{\it Left panel}: Transit light curves for a planet with 
$R_P = 2 R_{\rm Earth}$ orbiting an M5 star (dashed line: $P$), the same planet 
surrounded by a constant-absorptance mirror fleet extending to 3 planetary 
radii (solid line: $P+M$), and a planet without mirrors large enough to produce 
a light curve with the same depth of transit (dotted line: $LP$). 
{\it Right panel}: Difference between the mirror fleet transit light curve ($P+M$) 
and the one for the solitary large planet ($LP$), 
relative to the stellar intensity.
\label{fig:M5}}
\end{figure}

The top image in Figure \ref{fig:M5_img} displays our reference simulation:
a planet with a radius $R_P = 2 R_{\rm Earth}$, 
surrounded by a constant-absorptance fleet
of mirrors with $R_m = 3 R_P$, transiting in front of an M5 star.  The 
solid line in the left panel of Figure \ref{fig:M5} displays the simulated
light curve for the planet and its fleet of mirrors ($P+M$), with the
dashed line representing the planet-only transit curve ($P$).
A transit of the same depth as $P+M$ but caused by an isolated larger planet
($LP$, as in the bottom of Figure \ref{fig:M5_img}) is shown by the dotted line.

Containing many individual reflectors, the translucent mirror fleet
blocks some starlight (see Figure \ref{fig:M5_img}) and deepens the transit. 
This $P+M$ case can
be distinguished from that of an isolated larger planet ($LP$) in the timing
and shape of the transit ingress and egress light curve.  In the initial
phases of the transit, only the mirrors block starlight,
so the decrease occurs relatively gradually.  When the planet itself
enters transit, the light curve decreases more rapidly. Later in
the sequence, the dimming slows with the planet fully in front of the
star while the semi-transparent halo of mirrors continues to occlude
more starlight.  These effects are seen in the left-hand panel of
Figure \ref{fig:M5}, although the signature is complicated by
the presence of stellar limb-darkening. 
Note that once the mirrors are fully in 
front of the star, as depicted in Figure \ref{fig:M5_img}, the transit 
light curve becomes very similar to that of a larger isolated planet, 
with the small differences caused by variations in stellar brightness 
across the differing areas of occlusion. 
The right panel shows the difference of the light
curves for the planet plus mirrors transit and the larger planet transit,
highlighting the effects as the planet enters and exits its transit.

Although a fleet of mirrors orbiting an exoplanet affects both the
time-span and depth of a transit, it would not affect the
timing or depth of the planetary eclipse optical light curve. 
Once again, a difference in transit and eclipse time-span could suggest
the presence of mirrors to illuminate a planetary dark side, but the
same caveat regarding elliptical planetary orbits applies here as for 
the single large mirror. We assume the mirror fleet is 
designed for maximum efficiency, so all starlight is directed
onto the planet and none is re-directed towards us during the eclipse.
However, if the 
fleet does not include efficiency as a primary design constraint, 
we might observe a temporary {\em increase} in brightness just
before or after the eclipse phase due to spillover effects.

We anticipate no detectable effects due to thermal emission from
either a single large mirror or an armada of orbiting mirrors. We assume
the back sides are cold, as mirror reflection efficiency 
is high and no bright IR sources illuminate them from behind. Since 
light mirrors are very thin, the front and back are at nearly 
the same cold temperature.

\subsection{Effects of Varying Stellar Abundance or Spectral Type}\label{sect:LC_SP}

\begin{figure}[tbp]
\ifpdf
\includegraphics[width=0.99\columnwidth,keepaspectratio=true]{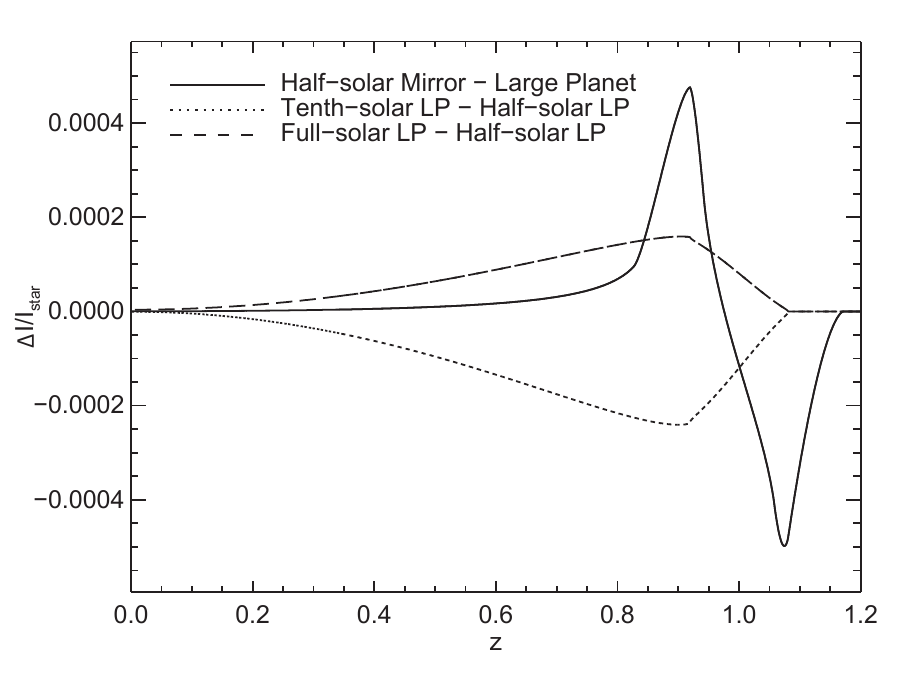}
\else
\includegraphics[width=0.99\columnwidth,keepaspectratio=true]{fig4.eps}
\fi
\caption{Impact on light curve of abundance-induced limb-darkening 
variations compared with that of a mirror fleet. The solid line displays 
the difference (relative to stellar intensity) between light curves 
produced by a planet surrounded by a mirror fleet and an isolated large 
planet, both transiting an identical M5 star with half-solar abundance.
The other two lines display the difference between the isolated large planet 
light curve for half-solar abundance and the light curves for isolated planets 
producing transits of the same depth for tenth-solar abundance (dotted) and
full-solar abundance (dashed).
\label{fig:M5_abund}}
\end{figure}

Since the mirror fleet mainly affects transit entry and
exit, we questioned whether differences 
in limb-darkening could exhibit the same signature. One
source of uncertainty in limb-darkening arises from abundance variations.
In runs for identical planets ($R_P = 2R_{\rm Earth}$, $R_m = 3 R_P$), 
varying the stellar abundance from half-solar to 
tenth-solar or to full-solar impacted the light curve
depth by 0.5\% to 4\%, depending on stellar type. This is of similar 
magnitude as the effects of a mirror fleet.  
We therefore created light curves for isolated planets transiting
M5 stars with tenth-solar and full-solar abundances, with sizes
generating transit depths equal to that for the reference simulation.
The solid line in Figure \ref{fig:M5_abund} shows the difference of the 
mirror plus planet and large planet light curves from Figure \ref{fig:M5}, 
while the dotted and dashed lines show the difference of the light
curves for the isolated planets for various stellar abundances. Different
stellar abundances cannot be mistaken for the effects of a transiting
planet surrounded by a fleet of mirrors, as limb-darkening variations
have a fundamentally different signature.

Although not displayed, we simulated transits of
planets orbiting the full range of stellar types displayed in Table
\ref{table:MS_stars}. Transit light curve details differ
somewhat for stars of earlier spectral type. In particular: 
(1) Differences in limb-darkening affect the detailed shape of the light curve
near entry and exit. (2) For larger stars, 
the same planet plus mirror fleet blocks a smaller fraction of light, 
resulting in a shallower transit, and the mirror fleet's effects last a
smaller fraction of the transit time-span.  Compared with an isolated 
planet transit of the same depth, the impact of the mirrors is of order
$10^{-4}I_{\rm star}$ for an M0 star, roughly a factor of 2 smaller
for a G2 star, and several times larger for an M5 star.
(3) Longer orbital periods, corresponding to the larger HZ orbits around
earlier-type stars, increase the time span of transits for these planets.
Despite these differences, simulating transits of other stars does not 
significantly change the bipolar signature of the mirror fleet 
in transit residuals. In addition, the maximum residuals 
remain approximately the same relative to transit depth for all stellar
types (e.g. $\sim$4-6\% for $R_P = 2R_{\rm Earth}$, $R_m = 3R_P$).

\subsection{Varying Mirror Fleet Extent, Planetary Size or Orbit Inclination}

\begin{figure}[tbp]
\ifbw{
\ifms{
\ifpdf
\plotone{fig5_BW.pdf}
\else
\plotone{fig5_BW.eps}
\fi
}
\ifpp{
\ifpdf
\includegraphics[width=0.99\columnwidth,keepaspectratio=true]{fig5_BW.pdf}
\else
\includegraphics[width=0.99\columnwidth,keepaspectratio=true]{fig5_BW.eps}
\fi
}
}
\ifcol{
\ifms{
\ifpdf
\plotone{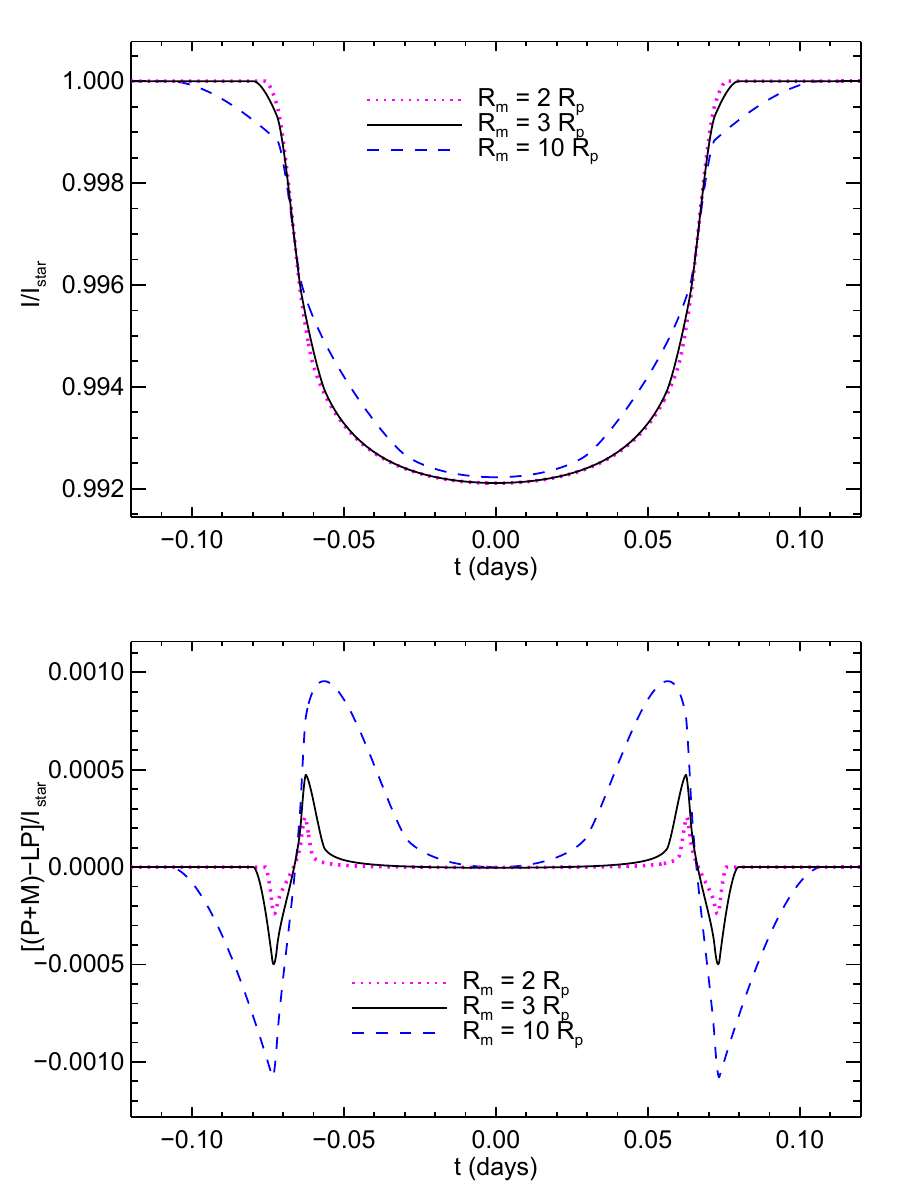}
\else
\plotone{fig5_col.eps}
\fi
}
\ifpp{
\ifpdf
\includegraphics[width=0.99\columnwidth,keepaspectratio=true]{fig5_col.pdf}
\else
\includegraphics[width=0.99\columnwidth,keepaspectratio=true]{fig5_col.eps}
\fi
}
}
\caption{{\it Top panel}: Transit light curves that result when a planet with 
$R_p = 2 R_{\rm Earth}$, located in the middle of the star's
HZ, passes in front of an M5 star. In all cases
the planet is surrounded by a constant-absorptance mirror fleet, with 
$R_m = 3R_P$ (solid), $R_m = 2 R_p$ (dotted), or $R_m = 10 R_p$ (dashed).
{\it Bottom panel}: Difference between the mirror fleet transit light curve ($P+M$) 
and the one for a solitary larger planet ($LP$) that would produce the
same depth of transit, relative to the stellar intensity, 
for the same situations.
\label{fig:M5_mr_diff}}
\end{figure}

The top panel of Figure \ref{fig:M5_mr_diff}
displays light curves for transits that are identical to the reference
simulation apart from the mirror fleet's extent. 
Larger fleets affect the light curve for a larger fraction of the transit
duration, with more dramatic departures from the isolated planet situation.
The dashed curve ($R_m = 10R_P$) has a slightly different transit depth
because 
$R_P/R_{\rm star} = 0.057$ so $R_m/R_{\rm star} = 0.57$; the 
mirror fleet extends across a significant fraction of the stellar surface,
which has different intensities at different locations due 
limb-darkening.
Because all mirror fleets have the same effective area, 
the larger fleet has a lower absorptance. So at $z = 0$
the smaller, more opaque systems block light primarily from the 
star's brighter central regions,  
while the largest, most transparent one transmits more of that light, 
instead blocking less intense outer portions of the star.
The bottom panel illustrates that, 
all else being equal, mirror fleets with large spatial extent would 
be more apparent in transit light curve abnormalities than those with 
smaller spatial extent, both temporally and in the magnitude of the effect
on transit residuals.

\begin{figure}[tbp]
\ifbw{
\ifms{
\ifpdf
\begin{center}
\includegraphics[width=0.75\columnwidth]{fig6_BW.pdf}
\end{center}
\else
\begin{center}
\includegraphics[width=0.75\columnwidth]{fig6_BW.eps}
\end{center}
\fi
}
\ifpp{
\ifpdf
\includegraphics[width=0.99\columnwidth]{fig6_BW.pdf}
\else
\includegraphics[width=0.99\columnwidth]{fig6_BW.eps}
\fi
}
}
\ifcol{
\ifms{
\ifpdf
\begin{center}
\includegraphics[width=0.75\columnwidth]{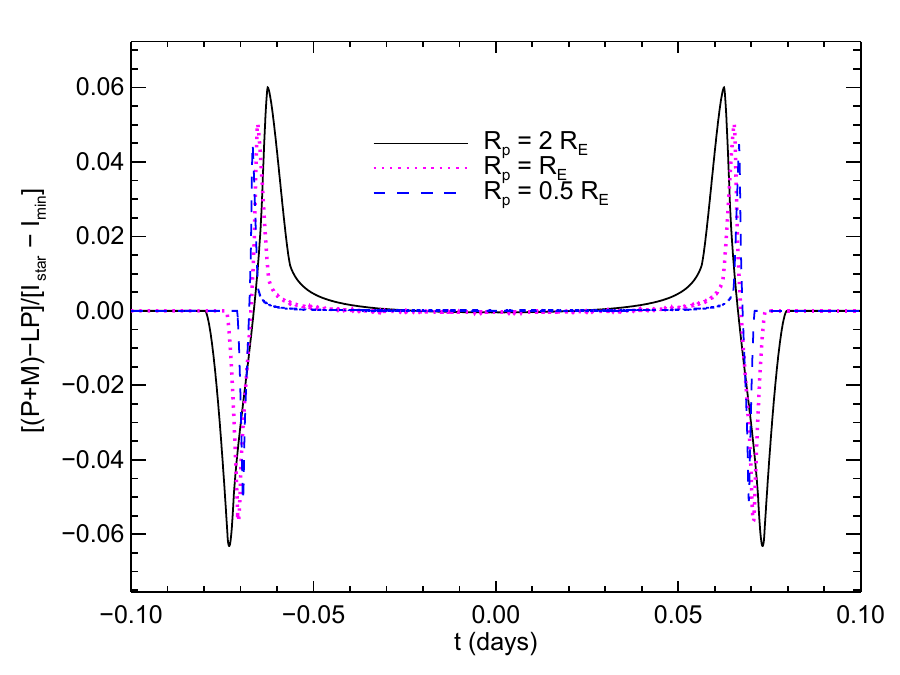}
\end{center}
\else
\begin{center}
\includegraphics[width=0.75\columnwidth]{fig6_col.eps}
\end{center}
\fi
}
\ifpp{
\ifpdf
\includegraphics[width=0.99\columnwidth]{fig6_col.pdf}
\else
\includegraphics[width=0.99\columnwidth]{fig6_col.eps}
\fi
}
}

\caption{Impact of a fleet of mirrors with $R_m = 3R_P$ on transits 
for planets of various sizes orbiting in the middle of an M5 star's HZ.
The lines show the difference between the light curve resulting from 
the planet plus mirror ($P + M$) transit and one of the same depth 
resulting from an isolated larger planet ($LP$) transit for planets
with $R_P/R_{\rm Earth} = 0.5$ (dashed), 1.0 (dotted), and 2.0 (solid).
\label{fig:M5_rp_diff}}
\end{figure}

For mirror fleets as in the reference model, we explored the effects
of varying planet size.  Figure \ref{fig:M5_rp_diff} displays the 
the difference between the $P+M$ and $LP$ light curves, 
relative to transit depth
$I_{\rm star} - I_{\rm min}$, for three sizes of terrestrial planets,
each orbiting at HZ$_{\rm mid}$ around an M5 star. 
Smaller planets result in transits of substantially shallower depth, 
however in all cases the mirror fleet has a significant impact 
(a few percent of the transit depth) on the entrance and exit light curves.
The duration of largest residuals is shorter for smaller planets.

\begin{figure}[tbp]
\ifbw{
\ifms{
\ifpdf
\plotone{fig7_BW.pdf}
\else
\plotone{fig7_BW.eps}
\fi
}
\ifpp{
\ifpdf
\includegraphics[width=0.99\columnwidth,keepaspectratio=true]{fig7_BW.pdf}
\else
\includegraphics[width=0.99\columnwidth,keepaspectratio=true]{fig7_BW.eps}
\fi
}
}
\ifcol{
\ifms{
\ifpdf
\plotone{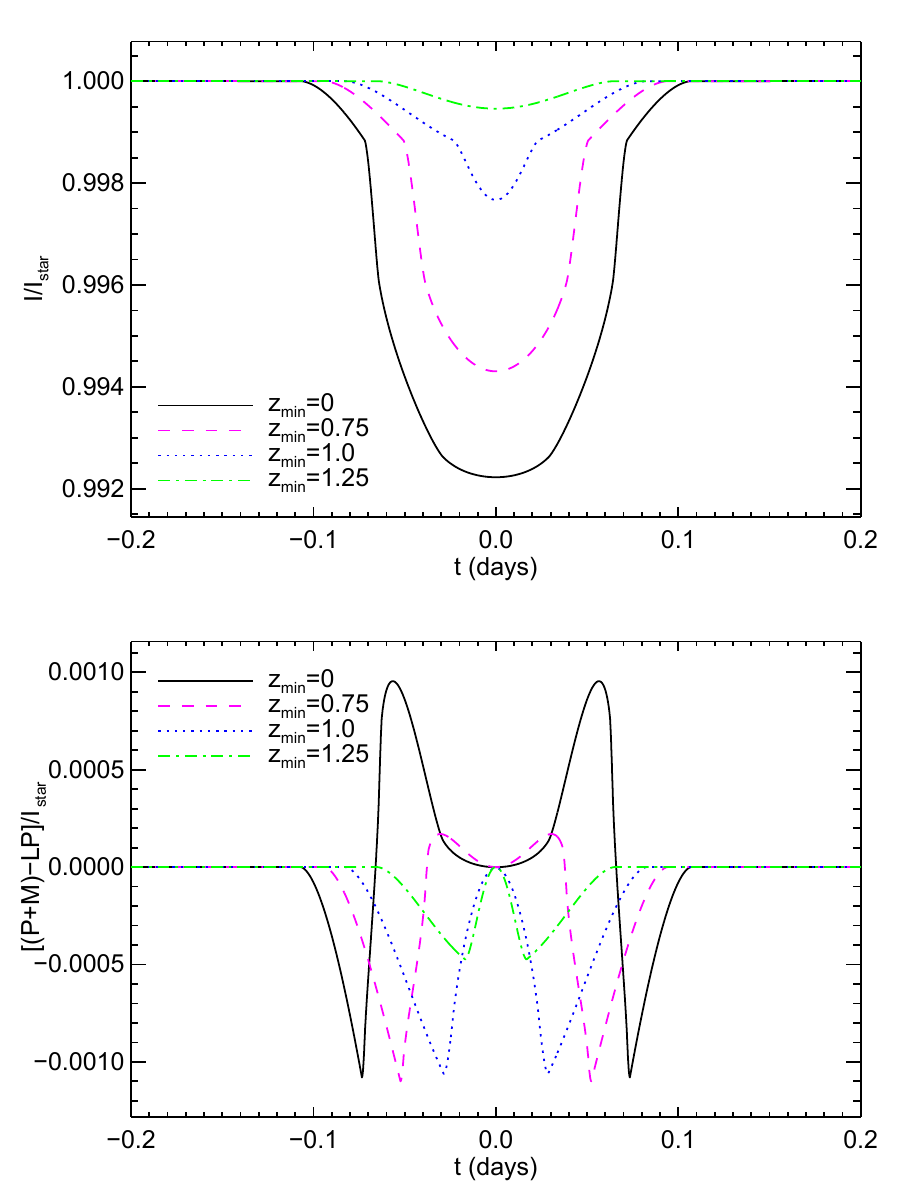}
\else
\plotone{fig7_col.eps}
\fi
}
\ifpp{
\ifpdf
\includegraphics[width=0.99\columnwidth,keepaspectratio=true]{fig7_col.pdf}
\else
\includegraphics[width=0.99\columnwidth,keepaspectratio=true]{fig7_col.eps}
\fi
}
}
\caption{{\it Top panel}: Transit light curves that result when a planet with 
$R_p = 2 R_{\rm Earth}$, located in the middle of the star's
HZ, passes in front of an M5 star. In all cases
the planet is surrounded by a constant-absorptance mirror fleet with 
$R_m = 10 R_P$. The four curves result from different values for the 
impact parameter:
$z_{\rm min} = 0$ (solid), $z_{\rm min} = 0.75$ (dashed), 
$z_{\rm min} = 1$ (dotted), and $z_{\rm min} = 1.25$ (dash-dot).
{\it Bottom panel}: Difference between the mirror fleet transit light curve ($P+M$) 
and the one for a solitary larger planet ($LP$) that would produce the
same depth of transit, 
relative to the stellar intensity, for the same situations.
\label{fig:M5_impact}}
\end{figure}

All previous results are for edge-on orbits, so the
minimum projected distance between star and planet centers is 
$z_{\rm min} = 0$.  The top panel of Figure \ref{fig:M5_impact} compares 
this situation (solid line) with orbits that are not
exactly edge-on ($z_{\rm min} \ne 0$). 
For these, a planet with $R_P = 2R_E$ and $R_m = 10 R_P$ is located at 
HZ$_{\rm mid}$ around an M5 star.
The dashed line planet transits at $z_{\rm min} = 0.75$, so part of the 
mirror fleet never occludes the star ($R_m/R_{\rm star} = 0.57$).  
In the dotted line case, 
the planet center crosses the edge of the star ($z_{\rm min} = 1$), and for
the dash-dot light curve ($z_{\rm min} = 1.25$) the planet
itself never crosses in front of the star, but the mirrors
still create a transit.  For each impact parameter $z_{\rm min}$,
the bottom panel of Figure \ref{fig:M5_impact} shows 
the difference of the $P+M$ and $LP$ light curves.
This figure reveals that as $z_{\rm min}$ increases, the mirror effects are
evident during a greater fraction of the transit time-scale.
As long as more than half of the planet crosses in front of the star, i.e.
$z_{\rm min} < 1$,  the transit residuals display a bipolar signature of 
the mirror fleet's presence. For $z_{\rm min} > 1$, the residuals are no 
longer bipolar, but the mirror fleet produces transits that last much 
longer than would be appropriate for the transit depth in the absence of 
mirrors, and the light curve minimum is not at transit center.  

\ \par
\ \par
\ \par
\subsection{Effects of Different Mirror Density Distributions}\label{sect:mirreff}

\begin{figure}[tbp]
\ifpdf
\plotone{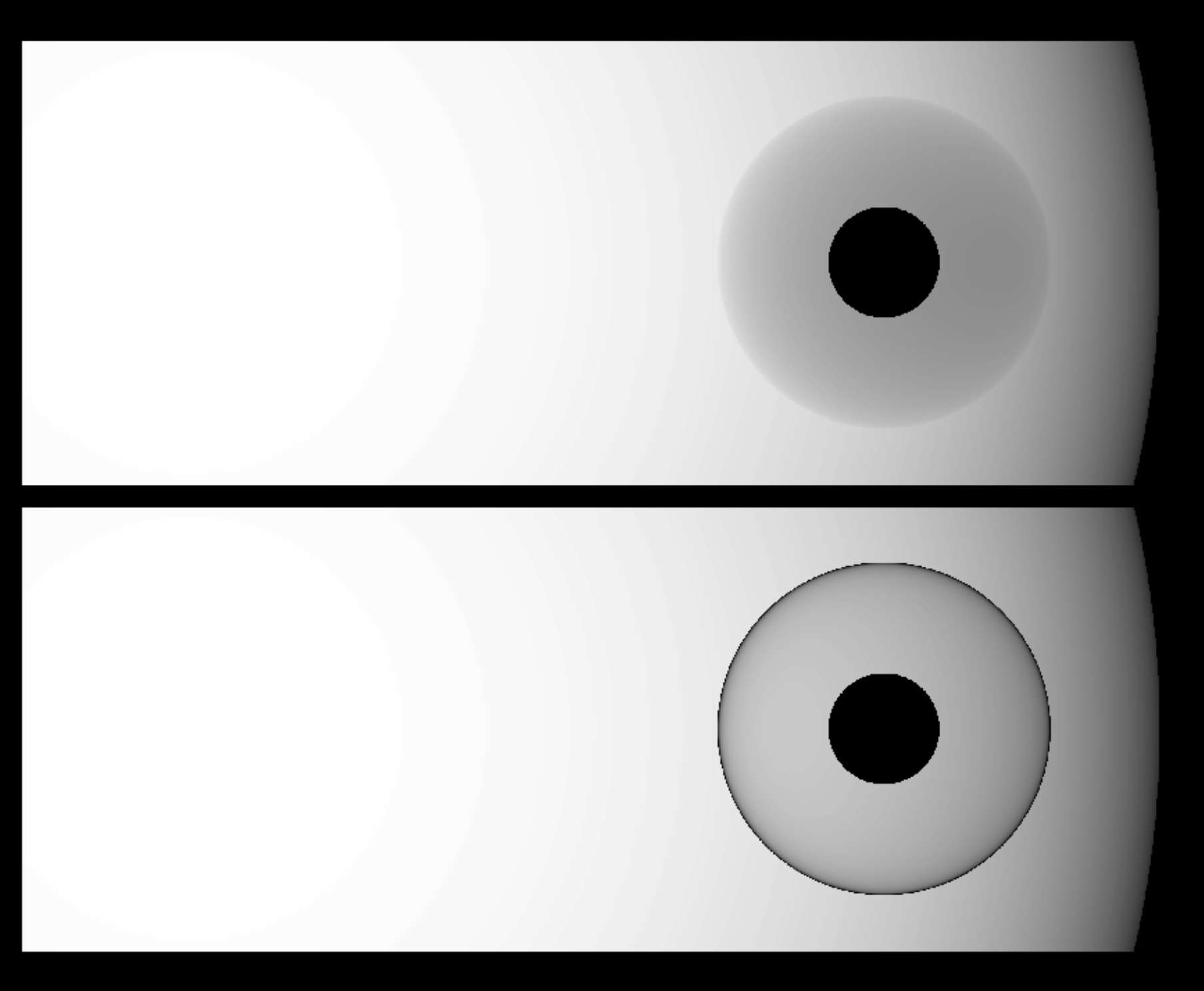}
\else
\plotone{fig8.eps}
\fi
\caption{{\it Top panel}: Simulated image for the same situation as Figure 
\ref{fig:M5_img} but for a planet surrounded by a constant-density fleet 
of mirrors in a thick shell extending from just above the planet's surface
(from 1.01$R_P$ to 3$R_P$).
{\it Bottom panel}: Same as above, but the mirrors are distributed in 
a thin shell extending from 2.99$R_P$ to 3$R_P$.
\label{fig:M5_shell_img}}
\end{figure}

Having explored the first-order effects of a fleet of mirrors, we
considered alternative transmittance models based
on actual distributions of steerable mirrors, which 
required our brute-force approach. 
The simplest such situation is with mirrors evenly distributed 
(density $\rho_{\rm mirr}$) within a spherical shell 
($R_i \le r \le R_o$) around
the planet. Although not necessarily realistic, this scenario lets us
explore the importance of our constant-absorptance assumption.
We calculated the absorptance at each projected radial distance based on the
projected density of the fleet:
\begin{equation} 
\begin{split}
\rho_{\rm proj} = 2 \rho_{\rm mirr} (\sqrt{R_o^2-r^2} - \sqrt{R_i^2-r^2}) \qquad (r \lt R_i) \\
\rho_{\rm proj} = 2 \rho_{\rm mirr} (\sqrt{R_o^2-r^2}) \qquad (R_i \le r \le R_o) \\
\end{split}
\end{equation}
The absorptance was again normalized so the total light intercepted 
equals that normally incident on the planet, 
while simultaneously ensuring absorptance $\alpha \le 1$. 
We considered two 
extremes: a thick uniform shell of mirrors extending outward from just 
above the planet's surface, and mirrors all at nearly the 
same altitude.  In the former case, the mirror fleet's collective 
absorptance is largest near the planet's surface (near $R_i$), 
gradually decreasing 
to zero at the fleet's outer extent. For the thin-shell case, the 
absorptance is small near the planet's surface, 
increasing with distance, then sharply peaked near the fleet's
altitude.  Figure \ref{fig:M5_shell_img} illustrates the extreme differences 
in absorptance profiles for the thick-shell (top panel) and thin-shell 
(bottom panel) cases.

\begin{figure}[tbp]
\ifbw{
\ifms{
\ifpdf
\begin{center}
\includegraphics[width=0.75\columnwidth,keepaspectratio=true]{fig9_BW.pdf}
\end{center}
\else
\begin{center}
\includegraphics[width=0.75\columnwidth,keepaspectratio=true]{fig9_BW.eps}
\end{center}
\fi
}
\ifpp{
\ifpdf
\includegraphics[width=0.99\columnwidth,keepaspectratio=true]{fig9_BW.pdf}
\else
\includegraphics[width=0.99\columnwidth,keepaspectratio=true]{fig9_BW.eps}
\fi
}
}
\ifcol{
\ifms{
\ifpdf
\begin{center}
\includegraphics[width=0.75\columnwidth,keepaspectratio=true]{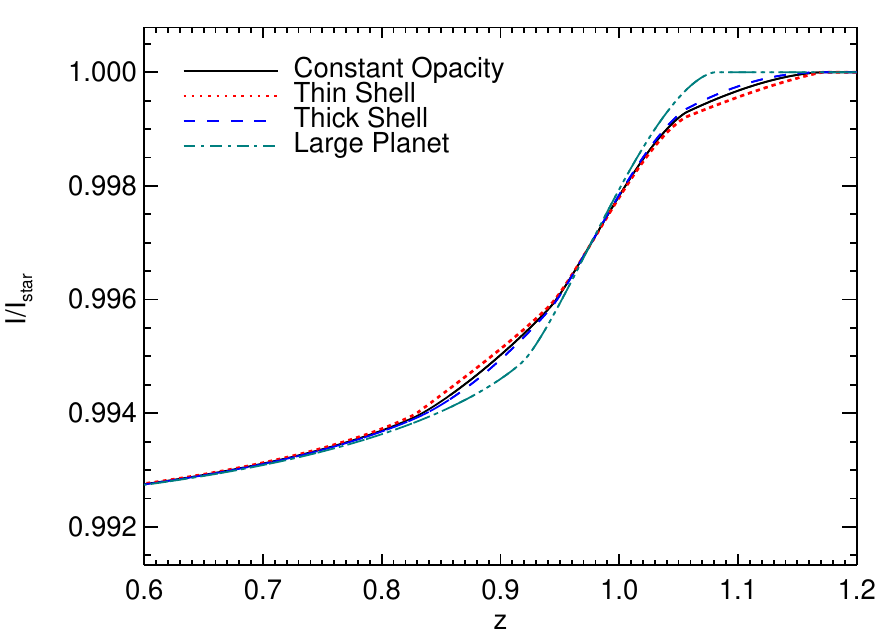}
\end{center}
\else
\begin{center}
\includegraphics[width=0.75\columnwidth,keepaspectratio=true]{fig9_col.eps}
\end{center}
\fi
}
\ifpp{
\ifpdf
\includegraphics[width=0.99\columnwidth,keepaspectratio=true]{fig9_col.pdf}
\else
\includegraphics[width=0.99\columnwidth,keepaspectratio=true]{fig9_col.eps}
\fi
}
}
\caption{Transit light curves for different absorptances. 
The solid line displays the light curve of a transit produced
by a planet with $R_p = 2 R_{\rm Earth}$, and a constant-absorptance fleet of
mirrors extending outward to $R_m = 3 R_p$, crossing in front of the 
center of an M5 star with half-solar abundance. The dotted and dashed lines
display the transit light curve produced by thick ($R_i = 1.01 R_p$) 
and thin ($R_i = 2.99 R_p$) constant-density shells of mirrors, respectively, 
each with an outer radius of $3 R_p$.  The dash-dot line is the light 
curve for an isolated planet large enough to produce a transit of the 
same depth. 
\label{fig:M5_multi}}
\end{figure}

These simulations illustrate the relative insensitivity of transit 
light curves to varying mirror-fleet absorptance profiles.
Figure \ref{fig:M5_multi} compares transit light curves for these 
two cases with previously discussed results.   
The constant-absorptance and large planet light curves are the same as those
shown in Figure \ref{fig:M5}. 
The dotted line results from a thin shell
of mirrors ($R_i = 2.99 R_p$, $R_m = 3 R_p$), and
the dashed line from a thick shell of mirrors extending from 
$R_i = 1.01 R_p$ to $R_m = 3 R_p$. The inner radii in both extremes were
chosen based on the resolution of our brute-force
simulations (planet radius = 100 pixels).  From this
figure, it is apparent that light curves for various absorptance distributions
are more similar to each other than they are to that for an isolated planet.
Our simple constant-absorptance model therefore captures the dominant
effects of this method of dark-side illumination.

\section{Prospects for Detection}\label{sect:detect}

We now consider the potential detectability of such
mirror installations, focusing 
on the predicted ($P+M$) light curve for an Earth-sized planet 
($R_P$ = $R_{\rm Earth}$) orbiting a G2 star at HZ$_{\rm in}$,
surrounded by a large ($R_m = 10 R_P$) constant-absorptance fleet of mirrors.
We chose this situation in part because
the only known civilization evolved on such a planet. 
Older M stars are very common, and their potentially habitable
terrestrial planets are likely in synchronous rotation. However they are 
fainter, so not necessarily better candidates when
seeking signatures of planetary-scale engineering.
Recall that a planet near HZ$_{\rm in}$ for a 
G2 star might be synchronously rotating at this time (see
Section \ref{subsect:HZ}), or that civilizations might choose to affect
the climate of bodies other than their planet of origin.  

\subsection{Detectability with {\it Kepler}}

Until now we have neglected the reality that a mirror fleet's effects 
will be partially suppressed by transit
fitting routines through variations in orbit inclination or stellar
limb-darkening.  To assess this, we re-sampled our $P+M$ model onto
\Kepler's short cadence (SC) timescale for 3 transits.  We used
the Transit Analysis Package (TAP, \citet{gazak2012}) to fit an isolated
planet model to the simulated light curve, restricting limb-darkening
parameters to a range appropriate for stars with effective temperatures within  750K of a G2 star.  
Because the fleet is translucent ($\alpha \sim0.01$ for $R_m
= 10 R_P$), predicted residuals ($\sim 10^{-5}I_{\rm star}$)
are substantially smaller than those
for the Jupiter-scale opaque structures of \citet{arnold05},
and thus too small to detect in \Kepler\ data.

To confirm this, we simulated ``best-case scenario" \Kepler\ data for our 
$P+M$ model and compared the residuals for the previously determined 
isolated planet model with the difference between the $P+M$ model and the 
isolated-planet TAP model determined above (hereafter TAP IPM).  
Given the relatively long orbital period ($\sim$ 237d) for
a planet orbiting a G2 star at HZ$_{\rm in}$ ($a = 0.75$ AU), at most 6 
transits of data are available. Kepler 10 is a relatively bright 
(Kp = 10.96) G2 star in the \Kepler\ sample, with instrumental and photon 
noise estimated by the \Kepler\ pipeline as $\sim$200 parts-per-million (ppm)
for short-cadence data.
We simulated 6 transits of SC data with Gaussian white noise at this noise 
level, although we note that \citet{batalha2011} report more 
long-cadence noise
($\sim$62 ppm) than the \Kepler\ pipeline estimate ($\sim$ 36 ppm).

\begin{figure}[tbp]
\ifbw{
\ifms{
\epsscale{0.66}
\ifpdf
\plotone{fig10_BW.pdf}
\else
\plotone{fig10_BW.eps}
\fi
}
\ifpp{
\ifpdf
\includegraphics[width=0.99\columnwidth,keepaspectratio=true]{fig10_BW.pdf}
\else
\includegraphics[width=0.99\columnwidth,keepaspectratio=true]{fig10_BW.eps}
\fi
}
}
\ifcol{
\ifms{
\epsscale{0.66}
\ifpdf
\plotone{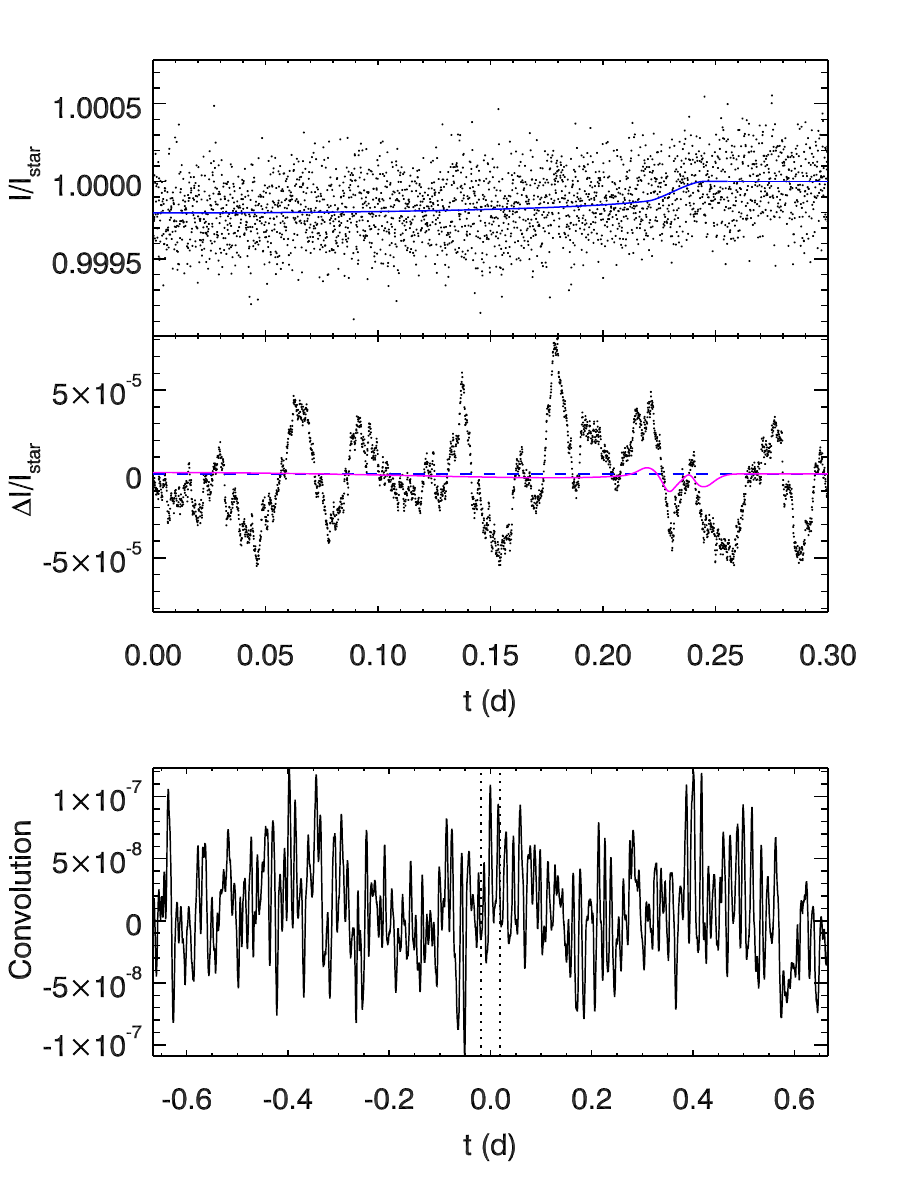}
\else
\plotone{fig10_color.eps}
\fi
}
\ifpp{
\ifpdf
\includegraphics[width=0.99\columnwidth,keepaspectratio=true]{fig10_color.pdf}
\else
\includegraphics[width=0.99\columnwidth,keepaspectratio=true]{fig10_color.eps}
\fi
}
}
\caption{{\it Top panel}: Simulated phase-folded data with 200 ppm noise 
and \Kepler\ SC sampling for 6 transits of a planet with 
$R_P = R_{\rm Earth}$ and $R_m = 10 R_P$, orbiting around a G2 star at 
0.75 AU. Only half of the transit is displayed.  The solid line is 
the best-fit isolated planet model produced by TAP analysis of the 
noiseless $P+M$ model.
{\it Middle panel}: Smoothed simulated residuals (dots) of the simulated 
data and TAP model shown in the top panel, and predicted residuals
(solid line) based on the difference between the isolated planet TAP 
model and the simulated $P+M$ model. The smoothing length for both was
determined by half of the time-scale $t_{1/2}$ (defined in the text).
{\it Bottom panel}: Convolution of the unsmoothed simulated and predicted
residuals. The vertical dotted lines indicate $\pm t_{1/2}$, 
and relate to the statistical analysis described in the text regarding
Figure \ref{fig:detect}.
\label{fig:kepdet}}
\end{figure}

The top panel of Figure \ref{fig:kepdet} displays the phase-folded
simulated data, where we retain the folded time scale for convenience in later 
analysis and t=0 represents the center of each transit.
For clarity, only half of the transit is displayed.
The superimposed solid line is the TAP IPM described above,
i.e. the one that best fits the noiseless $P+M$ model used to
simulated the data.
The residuals between our noisy data and the TAP IPM,
smoothed for visual clarity, are shown in the middle panel. They are 
overlaid (solid line)
by the smoothed difference between our $P+M$ model and the TAP IPM.
Clearly, noise in these data overwhelms any potential mirror fleet signal.
In the bottom panel, the convolution of the (unsmoothed)
simulated and predicted residuals confirms they are uncorrelated.

M stars are substantially dimmer at visible wavelengths, resulting in
noisier data. A typical \Kepler\ catalog star has a magnitude of
$\sim 14$, based on information retrieved from the Exoplanet
Data Explorer \citep{wright2011}, but the median M star 
is fainter than $Kp = 15$.  Kepler 138, hosting multiple terrestrial
planets \citep{kipping14b,rowe14}, is an unusually bright \Kepler\
M0-dwarf at $Kp = 12.93$, but still two magnitudes fainter than the G2
star simulated below.  Thus the expected
noise level would be $\sim 2.5$ times greater. Because of the 
shorter orbital period for a planet at HZ$_{\rm in}$ around an M0 star, 
up to 5 times more transits of \Kepler\ data could be available,
mitigating the additional noise.
However mirror fleets around these stars would still be
undetectable in \Kepler\ data.

\subsection{Future Prospects for Detection}

The James Webb Space Telescope ({\it JWST}) will have a collecting area
of 25 m$^2$, 35 times larger than \Kepler\ (0.708 m$^2$).
Using area, quantum efficiency, and filter transmission parameters 
provided in \citet{KIH09} and 
\url{http://ircamera.as.arizona.edu/nircam/features.html}, along with the 
Planck blackbody function for a G2 star, 
we computed the expected count rate of the NIRCAM F150W2 filter relative to 
the Kepler full bandpass. The {\it JWST}
F150W2 count rate will be $\sim 45$ times that for Kepler,
reducing noise levels to $\sim 200/6.7 = 30$ ppm.

M stars emit a larger fraction of their light at infrared wavelengths, 
so we used the effective temperatures and radii of Table 
\ref{table:MS_stars} to 
calculate the relative expected {\it JWST} photon rates for G2, M0, 
and M5 stars at equal distances.
In the F150W2 bandpass, we expect 6.7 and 53 times
more photons from a G2 star than for M0 and M5 stars, respectively. 
Data for M stars at similar
distances will be noisier, requiring more transits to achieve 
detections, so only the closest and brightest M stars will be 
feasible targets for {\it JWST}.


As before, we simulated data with short-cadence sampling for an Earth-like 
planet orbiting at a G2 star's HZ$_{\rm in}$ using our $P+M$ model, 
but at {\it JWST}'s expected noise level of 30 ppm.
We compared the residuals between the simulated data and the 
TAP IPM with the difference between the 
$P+M$ model (used to simulate the data) and the TAP IPM.
We did this comparison for 100 separate realizations of noise for each of
2,3,4,5, and 10 transits of JWST data. 
Although we would ideally base our
comparisons on isolated-planet TAP models fit to each simulated noise
realization, this simplified first-order analysis is sufficient 
to illustrate the key points regarding mirror fleet detectability.

\begin{figure}[tbp]
\ifbw{
\ifms{
\epsscale{0.7}
\ifpdf
\plotone{fig11_BW.pdf}
\else
\plotone{fig11_BW.eps}
\fi
}
\ifpp{
\ifpdf
\includegraphics[width=0.99\columnwidth,keepaspectratio=true]{fig11_BW.pdf}
\else
\includegraphics[width=0.99\columnwidth,keepaspectratio=true]{fig11_BW.eps}
\fi
}
}
\ifcol{
\ifms{
\epsscale{0.7}
\ifpdf
\plotone{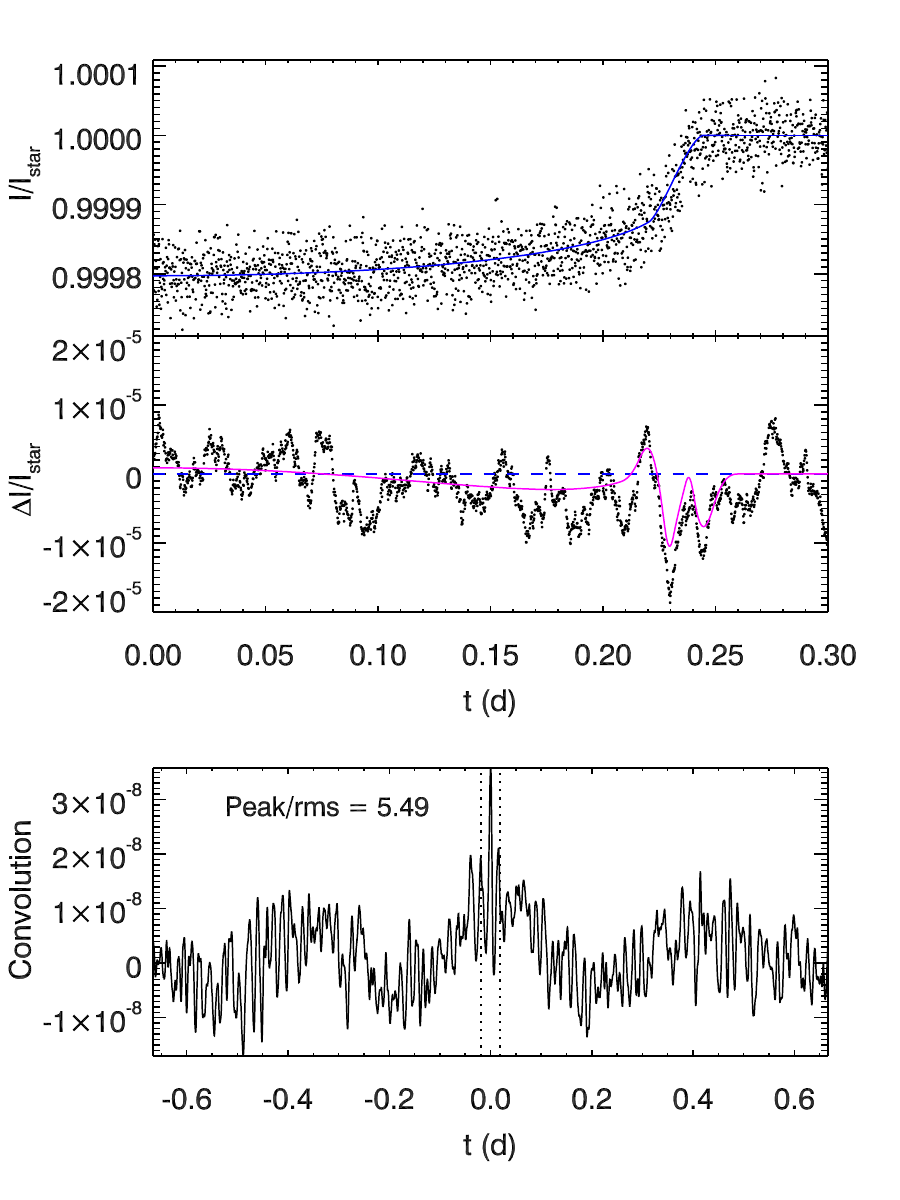}
\else
\plotone{fig11_col.eps}
\fi
}
\ifpp{
\ifpdf
\includegraphics[width=0.99\columnwidth,keepaspectratio=true]{fig11_col.pdf}
\else
\includegraphics[width=0.99\columnwidth,keepaspectratio=true]{fig11_col.eps}
\fi
}
}
\caption{Same as Figure \ref{fig:kepdet} but for 4 transits of simulated data 
with a noise level of 30 ppm, as anticipated for {\it JWST}. 
\label{fig:detect}}
\end{figure}

The top panel of Figure \ref{fig:detect} shows the phase-folded data
for a 4-transit noise realization, along with the TAP IPM.
For visual clarity, only half of the transit is displayed; however 
all analyses were performed using the full-transit simulated data.  
The smoothed residuals in the middle panel reveal the signature of
the mirror fleet; residuals of the simulated data and TAP IPM
(dots) are compared with predicted residuals (solid line; 
difference between simulated noiseless $P+M$ model and TAP IPM).
Both are smoothed by 
half of the time-scale $t_{1/2}$ (defined below) for visual clarity.
The convolution in the bottom panel demonstrates
the significant correlation between the (unsmoothed) simulated residuals and
the predicted model difference. Vertical dotted lines
delineate points satisfying 
$|\Delta t| \le t_{1/2}$, where $t_{1/2}$ is the time scale for
the isolated planet model light curve to decrease from maximum to half of 
the transit depth. The peak is 5.49$\sigma$ above
the convolution function's mean, with both mean and RMS ($\sigma$) 
determined by data outside the region delineated by $|t_{1/2}|$, i.e.
where we do not expect a significant signal.  We display this 
instance of noise because its convolution peak significance (5.49 $\sigma$) 
equals the median significance for the 100 separate 4-transit noise 
realizations.

In the 165 points with $|\Delta t| \le t_{1/2}$,
the probability of the convolution having a peak of at least this significance
occurring by chance is $4.4\times10^{-4}$\%. Of the
100 instances of 4 transits with white Gaussian noise of 30 ppm, the
convolution peak was at least 4$\sigma$ 94 times.  In our
5-transit simulations, 99/100 have convolution peaks
4$\sigma$ or higher, with a median of 6.10$\sigma$, and a 
probability of random occurrence no greater than $\sim 2$\% in all 
100 realizations.  Thus for a star like
Kepler-10, we anticipate successful detection of such a mirror
fleet with data for only a moderate number of transits. 

Could the effects of dark-side illumination be detectable
with fewer transits?  For the 100 separate 3-transit realizations,
the median significance of the convolution peak was 4.78$\sigma$, with
83/100 occurring at 4$\sigma$ or above. For the 2-transit realizations, the
median peak significance was 3.90$\sigma$, with only 44/100 reaching the
4$\sigma$ threshold.  We conclude that at this 30-ppm noise level,
spurious low-significance detections become increasingly likely 
with fewer than 4 transits of data.

\begin{figure}[tbp]
\ifbw{
\ifms{
\epsscale{0.7}
\ifpdf
\plotone{fig12_BW.pdf}
\else
\plotone{fig12_BW.eps}
\fi
}
\ifpp{
\ifpdf
\includegraphics[width=0.99\columnwidth,keepaspectratio=true]{fig12_BW.pdf}
\else
\includegraphics[width=0.99\columnwidth,keepaspectratio=true]{fig12_BW.eps}
\fi
}
}
\ifcol{
\ifms{
\epsscale{0.7}
\ifpdf
\plotone{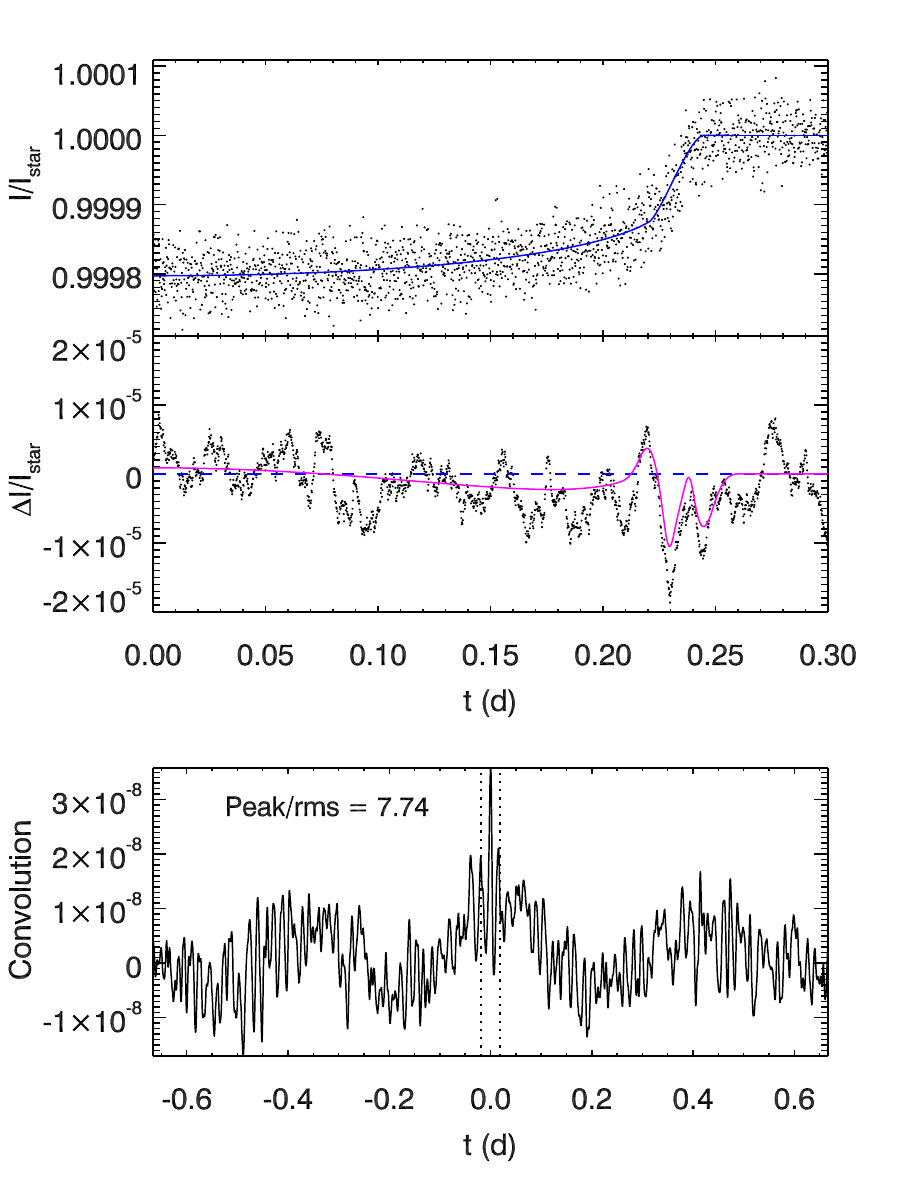}
\else
\plotone{fig12_col.eps}
\fi
}
\ifpp{
\ifpdf
\includegraphics[width=0.99\columnwidth,keepaspectratio=true]{fig12_col.pdf}
\else
\includegraphics[width=0.99\columnwidth,keepaspectratio=true]{fig12_col.eps}
\fi
}
}
\caption{Same as Figure \ref{fig:kepdet}, but for 10 transits of simulated data 
with a noise level of 30 ppm, as anticipated for {\it JWST}. 
\label{fig:jwst10}}
\end{figure}

An initial impression of the 4-transit residuals as shown
might suggest the isolated planet fit is adequate, so mirror fleet
signatures could go unnoticed without a correlation analysis similar to
that carried out above. However, for 10-transit simulations, the
convolution analysis results in a peak above 4$\sigma$ 100\% of the time,
with a probability of the weakest peak occurring by chance of $<$0.06\%. 
Figure \ref{fig:jwst10} displays the results of a 10-transit
simulation with a peak of median significance ($7.74\sigma$). 
In this situation, the simulated smoothed residuals do indicate that the
isolated planet model is insufficient, and the high significance of this
signature is made apparent by the strong peak in the 
convolution shown in the bottom panel. The probability of this peak 
occurring at random is only 2.7 $\times 10^{-10}$\%.

Residuals could be further suppressed if the isolated-planet
TAP models do not restrict the range of limb-darkening parameters to
values appropriate for similar stars. However, in the
majority of cases the stellar temperature and spectral type are known
sufficiently well to support the use of some restrictions.

The Transiting Exoplanet Survey Satellite ({\it TESS}), observing at
600-1000 nm, may identify transiting exoplanets for which targeted
{\it JWST} observations would be sufficient for our purpose 
(Ricker 2014\nocite{Ricker2014}, Ricker et al. 2014\nocite{RickerSPIE14}).
{\it TESS} will look at stars 30 to 100 times brighter than \Kepler\
($5 \lesssim\ m \lesssim\ 12$), across the full sky.  Primarily sensitive
to orbital periods up to 60 days, it can detect longer orbital periods
in regions of the sky coinciding with {\it JWST}'s continuous viewing
zone (CVZ).  It will study primarily F, G, and K stars, but also all
of the M dwarfs within 200 light years.  It is anticipated that 
{\it TESS} will identify $\sim$ 300 transiting super-Earths and tens of
Earth-sized planets, with $\sim 1/3$ of them in or near the CVZ.
{\it TESS} might well discover Earth-sized
exoplanets in the habitable zones of K or M stars bright enough for
detailed follow-up observations with {\it JWST} or future ground-based
30-m class telescopes. Since high-quality transit data on such planets
are of great interest for studying properties such as their atmospheres,
data adequate for the task of identifying or ruling out fleets of orbiting
mirrors will become available.

\subsection{Second-Order Effects \& Potential False Positives}\label{sect:2ndorder}

Since orbiting mirrors illuminate the planet's dark side, 
we explored the importance of planetary reflection of this light.
We assume the planetary contribution to 
photons received is constant, and that the amount of starlight reaching
the day side equals that reaching
the ``dark" side via mirrors. The planet to stellar flux ratio
is therefore of order 
$$\frac{F_P}{F_{\rm star}} = \left(\frac{R_P}{R_{\rm star}}\right)^2 \left(\frac{A}{2\pi}\right) \left(1 - \frac{\alpha}{2}\right)$$
The first factor represents the fraction of total starlight
intercepted by the cross-sectional area of the planet. 
$A$ is the planet's albedo, i.e.  the fraction of starlight 
that is scattered, with division by $2\pi$ because it scatters
into $2\pi$ sr.  The final factor accounts for dimming of 
scattered planetary light by the mirror fleet; absorptance is reduced
because only half the mirrors lie on our side of the planet.
Greater effects occur for planets that are large relative to the stellar size.  
For small M8 stars, a planet with $R_P = 2 R_{\rm Earth}$ 
has radius $R_P = 0.14 R_{\rm M8}$.  For a mirror fleet extending 
to $R_m = 3 R_P$, the absorptance (given our assumptions) is $\alpha = 0.125$. 
Assuming a planetary albedo comparable to Earth's ($A = 0.3$),
this contribution alters the observed transit depth 
by at most 0.089\%, significantly less than the 4-6\% due to mirror fleets
(see end of Section \ref{sect:LC_SP}). Thus scattered light from
the now-illuminated dark side produces only a small effect on
a transit light curve.  

Since {\it JWST} and other future telescopes
will observe primarily in the infrared, we must consider 
thermal radiation from the now-warmed dark side.  
The planet to stellar flux ratio at any particular wavelength is
$$\frac{F_P(\lambda)}{F_{\rm star}(\lambda)} = 
\frac{\epsilon B_{\lambda}(T_P)}{B_{\lambda}(T_{\rm star})}
\left(\frac{R_P}{R_{\rm star}}\right)^2 $$
where $B_{\lambda}(T)$ is the Planck thermal blackbody radiation,
and $\epsilon$ is the planet's emissivity. An upper limit
is achieved by assuming $\epsilon = 1$ and choosing the coolest,
smallest star (M8).  For an M8 star, $R_P/R_{\rm star} = 0.14$ and 
$T_{\rm star} = 2660$K; we assume a habitable planet with $T_P = 300$K. 
In the {\it JWST} bandpass (0.6$\mu$m-28.5$\mu$m), the flux ratio 
reaches a maximum of 0.09\% at a wavelength of 28.5$\mu$m. At the 
wavelength of strongest planetary emission ($\sim$ 10$\mu$m), 
the flux ratio is only 0.01\%. The change in transit depth is 
significantly smaller than the 4-6\% effects of orbiting mirrors.
Even if $T_P$ = 1000K, planetary thermal radiation
only alters the transit depth by 0.7\%.

In general, any potential detection of intelligence must be confirmed 
using an entirely different method. 
We must consider whether natural phenomena could mimic signs of intelligence.  
In particular, a face-on
ring system is extremely similar to the translucent projected annulus of
our putative fleet of mirrors. Several groups have modeled transit 
detectability of ring systems \citep{bf04,ohta09, tusnski11}, 
generally focusing on gas giant rings.  No known terrestrial
planet has a ring system, and we question the stability of such a
system around a terrestrial planet in tidally-induced
synchronous rotation, particularly since \citet{ohta09} found that rings 
around all close-in planets are likely to be short-lived and nearly
edge-on, rendering them more difficult to detect. 
More importantly, ring particles are likely warmer
than the mirrors, emitting significantly in the IR thus reducing the 
IR transit depth. Comparison of IR 
and optical light curves will therefore help discriminate between
natural and engineered systems, given sufficient sensitivity.
Ring systems may also be distinguished from mirror fleets
by the effects of forward scattering of light by ring particles, causing
flux to brighten just before and after planetary transit \citep{bf04}.

Planets in an M star's HZ 
might experience atmospheric erosion, due to strong stellar activity and weak
planetary magnetic moments (e.g. \nocite{Lammer07}Lammer et al.
2007 and discussions in \nocite{Scalo07}Scalo et al. 2007 and
\nocite{Tarter07}Tarter et al. 2007).  The 
excess absorption around the planet could produce a similar
transit light-curve signature; however mirrors are equally opaque at all
visible wavelengths, while atmospheric transmission
exhibits wavelength-dependent features.  Extensive optical/UV
transit spectroscopy observations of atmospheric escape from hot
jupiters HD 209458b reveal that escaping material contains
elements such as H, C, O, and Na \citep[see for example][]{,ehrenreich11,
linsky10,charbonneau02}, effects not expected for engineered structures.
Lyman-$\alpha$ transit observations of HD 189733b show
temporal variability in the evaporating atmosphere \citep{temporal12},
whereas a stable mirror fleet will produce no such fluctuations.
As a related example, \citet{rappaport12} explain transit depth variability 
for KIC 12557548 b with a trailing cloud of
material escaping a disintegrating super-mercury; this
phase-folded transit light curve exhibits substantial asymmetries.
Here transit depth should depend on wavelength, although 
multi-wavelength studies have not yet detected this \citep{croll14}.
In detailed modeling by 
\citet{brogi12}, a pre-transit flux excess is due
to forward scattering by dust, another feature not anticipated
with a mirror fleet designed for optimal efficiency. 

\citet{kipping14a} finds that bipolar residual signatures (somewhat different
in functional form from those discussed in Section \ref{sect:sims}) can result 
from uncorrected low-amplitude transit timing variations or 
from uncorrected variations in transit duration.  In principle, through trial
and error it should be possible to offset variations in either transit 
timing or transit duration and sharpen the transit entry and exit for the 
folded light curve, making residuals consistent with those for an isolated 
planet.  However if the bipolar residuals are due to a mirror fleet, they
will not be improved by any small 
adjustments to the timing or duration of each eclipse. Each subsesquent transit
would make the difference more pronounced.
Substantial technological improvements should eventually
yield sensitivity sufficient for detailed analysis of individual
transit light curves, in which case the mirror fleet would be clearly
distinguishable from these effects. 

\section{Conclusions}

An orbiting fleet of mirrors poses significant but possibly
not insurmountable environmental, economic, and engineering challenges,
so  could plausibly be utilized to redirect starlight onto a planet, 
perhaps to illuminate its dark side.
Differences in the time-span of planetary transits compared
with planetary eclipses could be one indication of the presence of such
structures, which would also affect the transit light curve.

We modeled transit light curves resulting from such fleets;
compared with an isolated-planet transit of the same depth, the transit
with orbiting mirrors begins earlier and reaches
its minimum more gradually, with differences most apparent near 
transit entry and exit.  The characteristic signature is not plausibly 
reproduced by alterations in limb-darkening due to abundance variations.
Regardless of stellar type, the difference is typically 4-6\% 
of the transit depth for mirrors extending 3 planetary radii beyond 
a planet with $R_P = 2 R_{\rm Earth}$. 
Alterations to planetary scattered light and thermal emission
caused by the mirrors are negligible compared to the mirrors' effects on
the light curve.
For smaller stars, mirror fleets alter transit depth by a 
larger fraction of the total starlight.  The difference is of order
$10^{-4}I_{\rm star}$ for an M0 star, roughly a factor of 2 smaller
for a G2 star, and several times larger for an M5 star.
Effects on light curves are most apparent for large planets
and mirror fleets of large radial extent.  
For orbital inclinations $|i|$ less than 90$\arcdeg$ ($z_{\rm min} \ne 0$),
mirror effects are evident during a greater fraction of the transit
time-scale, and produce inconsistencies between transit depth and time.
Variations in mirror fleet distribution produce very small 
alterations in the light curve, so a constant-absorptance model 
sufficiently captures the main features of such an engineering structure.

The mirror installations considered here are undetectable in 
the existing \Kepler\ data. However, with appropriate analysis, they
could be revealed in {\it JWST} observations, were they present.
Natural systems, such as rings or an evaporating atmosphere, 
might produce light curves with a similar signature, but the effects
will vary with wavelength quite differently than the mirror fleet case.
With sufficient follow-up observations and sensitivity, a mirror fleet
should be distinguishable from these effects.

These models likely do not accurately represent 
a realistic fleet of orbiting mirrors each adjusting its
attitude to maintain a desired illumination pattern. We anticipate this
effect is of similar scale to differences in mirror fleet geometry.
Future work will include detailed modeling of the
absorptance / transmittance of plausible mirror fleets, incorporating
consideration of the optimal range of mirror orbit size and eccentricity
to provide appropriate illumination, and variations of mirror angle
throughout the orbit. A similar effort needs to be made on whether stable
mirror orbits are achievable considering tidal effects of the star and radiative
effects of the illumination on the thin mirrors, as well as the feasibility of
propulsion systems capable of sustaining orbits under these conditions.

\acknowledgements
This research has made use of the Exoplanet Orbit Database and Exoplanet 
Data Explorer at \href{http://exoplanets.org}{exoplanets.org} and NASA's Astrophysics Data System.
During this work EJK was supported in part by NASA grant 
NNX09AN69G and donations from the Friends of SETI@home
(\href{http://setiathome.berkeley.edu}{setiathome.berkeley.edu}).  EJK would like to thank SMS for allowing him the
honor of first authorship for conceiving the original idea and describing
the basic analysis technique, despite the fact that SMS executed most of 
the work and did most of the writing.  Additional thanks to 
Jean-Paul Keulen for finding the error in our calculation of the launch cost of 
a mirror fleet.


\begin{thebibliography}{}

\bibitem[Arnold(2005)]{arnold05} Arnold, L.~F.~A.\ 2005, \apj, 
627, 534 
	\ifpp{\doi{10.1086/430437}}

\bibitem[Barnes \& Fortney(2004)]{bf04} Barnes, J.~W., \& Fortney, 
J.~J.\ 2004, \apj, 616, 1193
	\ifpp{\doi{10.1086/425067}}

\bibitem[Barnes et al.(2008)]{barnes08} Barnes, R., Raymond, S.~N., Jackson, 
B., \& Greenberg, R.\ 2008, Astrobiology, 8, 557
	\ifpp{\doi{10.1089/ast.2007.0204}}

\bibitem[Barnes et al.(2013)]{barnes13} Barnes, R., Mullins, K., Goldblatt, C.,
et al.\ 2013, Astrobiology, 13, 225
	\ifpp{\doi{10.1089/ast.2012.0851}}

\bibitem[Batalha et al.(2011)]{batalha2011} Batalha, N.~M, Borucki, W.~J.,
Bryson, S.~T., et al.\ 2011, \apj, 729, 27
	\ifpp{\doi{10.1088/0004-637X/729/1/27}}

\bibitem[Bowyer et al.(1997)]{bowyer97} Bowyer, S., Werthimer, 
D., Donnelly, C., et al.\ 1997, IAU Colloq.~161: Astronomical and 
Biochemical Origins and the Search for Life in the Universe, 667 
	\ifpp{\ads{1997abos.conf..667B}}

\bibitem[Brogi et al.(2012)]{brogi12} Brogi, M., Keller, C.~U., 
de Juan Ovelar, M., et al.\ 2012, \aap, 545, L5 
	\ifpp{\doi{10.1051/0004-6361/201219762}}
	
\bibitem[Bryant et al.(2014)]{bryant14} Bryant, R.~G., Seaman, S.~T., Wilkie,
W.~K., Miyauchi, M., \& Working, D.~C. 2014, 
in 3rd International Symposium on Solar Sailing, ed. M. Macdonald 
(Berlin Heidelberg: Springer-Verlag), 525
	\ifpp{\doi{10.1007/978-3-642-34907-2_33}}


\bibitem[Carrigan(2009)]{carrigan09} Carrigan, R.~A., Jr.\ 2009, 
  \apj, 698, 2075 
	\ifpp{\doi{10.1088/0004-637X/698/2/2075}}

\bibitem[Carrigan(2012)]{carrigan12} Carrigan, R.~A. Jr.\ 2012, Acta 
Astronautica, 78, 121 
	\ifpp{\doi{10.1016/j.actaastro.2011.12.002}}

\bibitem[Charbonneau et al.(2002)]{charbonneau02} Charbonneau, D., 
Brown, T.~M., Noyes, R.~W., \& Gilliland, R.~L.\ 2002, \apj, 568, 377 
	\ifpp{\doi{10.1086/338770}}

\bibitem[Claret \& Bloemen(2011)]{ClaretBloemen11} Claret, A., \& Bloemen, S.
2011, \aap, 529, A75 
	\ifpp{\doi{10.1051/0004-6361/201116451}}

\bibitem[Croll et al.(2014)]{croll14} Croll, B., Rappaport, S., 
DeVore, J., et al.\ 2014, \apj, 786, 100 
	\ifpp{\doi{10.1088/0004-637X/786/2/100}}

\bibitem[Darwin(1880)]{Darwin1880} Darwin, G.~H.\ 1880, Royal 
Society of London Philosophical Transactions Series I, 171, 713 
	\ifpp{\ads{1880RSPT..171..713D}}

\bibitem[Dressing \& Charbonneau(2013)]{dc2013} Dressing, C.~D., 
\& Charbonneau, D.\ 2013, \apj, 767, 95 
	\ifpp{\doi{10.1088/0004-637X/767/1/95}}

\bibitem[Dyson(1960)]{dyson60} Dyson, F.~J.\ 1960, \science, 131, 1667 
	\ifpp{\doi{10.1126/science.131.3414.1667}}

\bibitem[Eastman et al.(2013)]{eastman13} Eastman, J., Gaudi, 
B.~S., \& Agol, E.\ 2013, \pasp, 125, 83 
	\ifpp{\doi{10.1086/669497}}


\bibitem[Ehrenreich(2011)]{ehrenreich11} Ehrenreich, D.\ 2011, 
European Physical Journal Web of Conferences, 11, 03006 
	\ifpp{\doi{10.1051/epjconf/20101103006}}

\bibitem[Engle et al.(2009)]{engle09} Engle, S.~G., Guinan, E.~F.,
\& Mizusawa, T.\ 2009, 
American Institute of Physics Conference Series, 1135, 221 
	\ifpp{\doi{10.1063/1.3154054}}

\bibitem[Forget \& Leconte(2013)]{ForgetLeconte2013} Forget, F., \& Leconte, J.\
2014, Royal Soc. London Phil. Trans. Ser. A, 372, 30084
	\ifpp{\doi{10.1098/rsta.2013.0084}} 

\bibitem[Freitas(1983)]{freitas83} Freitas, R.~A., Jr.\ 1983, 
Journal of the British Interplanetary Society, 36, 501 
	\ifpp{\ads{1983JBIS...36..501F}}

\bibitem[Gaidos(2013)]{gaidos2013} Gaidos, E.\ 2013, \apj, 770, 90 
	\ifpp{\doi{10.1088/0004-637X/770/2/90}}

\bibitem[Gazak et al.(2012)]{gazak2012} Gazak, J.~Z., Johnson, J.~A., 
Tonry, J., et al.\ 2012, Advances in Astronomy, 2012, ID 697967
	\ifpp{\doi{10.1155/2012/697967}}

\bibitem[Grenfell et al.(2012)]{grenfell2012} Grenfell, J.~L., 
Grie{\ss}meier, J.-M., von Paris, P., et al.\ 2012, Astrobiology, 12, 1109 
	\ifpp{\doi{10.1089/ast.2011.0682}}

\bibitem[Griffith et al.(2015)]{griffith15} Griffith, R.~L., 
Wright, J.~T., Maldonado, J., et al.\ 2015, \apjs, 217, 25 
        \ifpp{\doi{10.1088/0067-0049/217/2/25}}

\bibitem[Henry et al.(2006)]{henry06} Henry, T.~J., Jao, W.-C., 
Subasavage, J.~P., et al.\ 2006, \aj, 132, 2360 
	\ifpp{\doi{10.1086/508233}}

\bibitem[Howard et al.(2007)]{howard07} Howard, A., Horowitz, 
P., Mead, C., et al.\ 2007, Acta Astronautica, 61, 78 
	\ifpp{\doi{10.1016/j.actaastro.2007.01.038}}

\bibitem[Hut(1981)]{Hut1981} Hut, P.\ 1981, \aap, 99, 126 
	\ifpp{\ads{1981A&A....99..126H}}

\bibitem[Jugaku \& Nishimura(2004)]{JugakuNish04} Jugaku, J., \& Nishimura, S.\ 2004, Bioastronomy 2002: Life Among the Stars, 213, 437 
	\ifpp{\ads{2004IAUS..213..437J}}

\bibitem[Kardashev(1964)]{kardashev64} Kardashev, N.~S.\ 1964, 
\sovast, 8, 217 
	\ifpp{\ads{1964SvA.....8..217K}}

\bibitem[Kasting et al.(1993)]{kasting93} Kasting, J.~F., 
Whitmire, D.~P., \& Reynolds, R.~T.\ 1993, \icarus, 101, 108 
	\ifpp{\doi{10.1006/icar.1993.1010}}

\bibitem[Kipping (2014)]{kipping14a} Kipping, D.~M. 2014, \mnras, 440, 2164
	\ifpp{\doi{10.1093/mnras/stu318}}

\bibitem[Kipping et al.(2014)]{kipping14b} Kipping, D.~M., 
Nesvorn{\'y}, D., Buchhave, L.~A., et al.\ 2014, \apj, 784, 28 
	\ifpp{\doi{10.1088/0004-637X/784/1/28}}

\bibitem[Kopparapu(2013)]{kopparapu2013} Kopparapu, R.~K.\ 2013, 
\apjl, 767, L8 
	\ifpp{\doi{10.1088/2041-8205/767/1/L8}}

\bibitem[Kopparapu et al.(2013)]{kopparapuetal13} Kopparapu, R.~K., 
Ramirez, R., Kasting, J.~F., et al.\ 2013, \apj, 765, 131 
	\ifpp{\doi{10.1088/0004-637X/765/2/131}, Erratum \doi{10.1088/0004-637X/770/1/82}}

\bibitem[Korpela et al.(2011)]{korpela11} Korpela, E.~J., 
Anderson, D.~P., Bankay, R., et al.\ 2011, \procspie, 8152, ID 815212
	\ifpp{\doi{10.1117/12.894066}}

\bibitem[Lammer et al.(2007)]{Lammer07} Lammer, H., 
Lichtenegger, H.~I.~M., Kulikov, Y.~N., et al.\ 2007, Astrobiology, 7, 185 
	\ifpp{\doi{10.1089/ast.2006.0128}}

\bibitem[Lecavelier des Etangs et al.(2012)]{temporal12} Lecavelier des Etangs, 
A., Bourrier, V., Wheatley, P.~J., et al.\ 2012, \aap, 543, L4 
	\ifpp{\doi{10.1051/0004-6361/201219363}}

\bibitem[Leconte et al.(2010)]{Leconte2010} Leconte, J., Chabrier, G., Baraffe, I., \& Levrard, B.\ 2010, \aap, 516, A64 
	\ifpp{\doi{10.1051/0004-6361/201014337}}

\bibitem[Lima et al.(2011)]{lima11} Lima, M.~D., Fang, S., Lepr{\'o}, X., 
et al.\ 2011, \science, 331, 51 
	\ifpp{\doi{10.1126/science.1195912}}

\bibitem[Linsky et al.(2010)]{linsky10} Linsky, J.~L., Yang, H., 
France, K., et al.\ 2010, \apj, 717, 1291 
	\ifpp{\doi{10.1088/0004-637X/717/2/1291}}

\bibitem[Mandel \& Agol(2002)]{MA02} Mandel, K., \& Agol, E.\ 2002, \apjl, 580, L171 
	\ifpp{\doi{10.1086/345520}}

\bibitem[Marcy et al.(2014)]{marcy14} Marcy, G.~W., Isaacson, 
H., Howard, A.~W., et al.\ 2014, \apjs, 210, 20 
	\ifpp{\doi{10.1088/0067-0049/210/2/20}}

\bibitem[Marcy et al.(2013)]{Marcy13_dyson}
Marcy, G., Howard, A., \& Johnson, J.\ 2013, 
``Discovery of Earth-like Planets and Signals from Intelligent Life,"  
Templeton Foundation New Frontiers in Astronomy and Cosmology Grant.

\bibitem[Mulders et al.(2014)]{mulders2014} Mulders, G.D., Ciesla, F., 
Pascucci, I., \& Apai, D. 2014, {/it The Water Content of Exo-earths in the 
Habitable Zone}, Search for Life Beyond the Solar System: Exoplanets,
Biosignatures \& Instruments
	\ifpp{\ads{2014ebi..conf.2.21M}}

\bibitem[Neron de Surgy \& Laskar(1997)]{nds_las1997} Neron de Surgy, O., \& Laskar, J.\ 1997, \aap, 318, 975 
	\ifpp{\ads{1997A&A...318..975N}}

\bibitem[Ohta et al.(2009)]{ohta09} Ohta, Y., Taruya, A., \& Suto, Y.\ 2009,
\apj, 690, 1
	\ifpp{\doi{10.1088/0004-637X/690/1/1}}

\bibitem[Petigura et al.(2013)]{petigura2013} Petigura, E.~A., 
Howard, A.~W., \& Marcy, G.~W.\ 2013, 
Proceedings of the National Academy of Science, 110, 19273 
	\ifpp{\ads{2013PNAS..11019273P}}

\bibitem[Pierrehumbert \& Gaidos(2011)]{PandG2011} Pierrehumbert, R., 
\& Gaidos, E.\ 2011, \apjl, 734, L13 
	\ifpp{\doi{10.1088/2041-8205/734/1/L13}}

\bibitem[Rappaport et al.(2012)]{rappaport12} Rappaport, S., 
Levine, A., Chiang, E., et al.\ 2012, \apj, 752, 1 
	\ifpp{\doi{10.1088/0004-637X/752/1/1}}

\bibitem[Ricker(2014)]{Ricker2014} Ricker, G.~R.\ 2014, 
{\it The Transiting Exoplanet Survey Satellite Mission}, 
Search for Life Beyond the Solar System: Exoplanets,
Biosignatures \& Instruments
        \ifpp{\ads{2014ebi..conf.3.10R}}

\bibitem[Ricker et al.(2014)]{RickerSPIE14} Ricker, G.~R., Winn, 
J.~N., Vanderspek, R., et al.\ 2014, SPIE J. Ast. Tel. Inst. \& Sys., accepted
\arxiv{1406.0151}

\bibitem[Rowe et al.(2014)]{rowe14} Rowe, J.~F., Bryson, 
S.~T., Marcy, G.~W., et al.\ 2014, \apj, 784, 45 
	\ifpp{\doi{10.1088/0004-637X/784/1/45}}

\bibitem[Sagan \& Drake(1975)]{sagan75}Sagan, C. \& Drake, F.\ 1975 \sciam, 232,
80
	\ifpp{\ads{1975SciAm.232...80S}}

\bibitem[Scalo et al.(2007)]{Scalo07} Scalo, J., Kaltenegger, 
L., Segura, A.~G., et al.\ 2007, Astrobiology, 7, 85 
	\ifpp{\doi{10.1089/ast.2006.0125}}, 

\bibitem[Shkolnik \& Barman(2014)]{shkolnik2014} Shkolnik, E., \& Barman, T.\ 2014, 
{\it HAZMAT I: The Evolution of X-ray Far-UV and Near-UV 
Emission from Early M Stars},
Search for Life Beyond the Solar System: Exoplanets,
Biosignatures \& Instruments
	\ifpp{\ads{2014ebi..conf..2.3S}}

\bibitem[Tarter(2001)]{tarter01}Tarter, J.\ 2001, \araa, 39, 511 
	\ifpp{\doi{10.1146/annurev.astro.39.1.511}}

\bibitem[Tarter et al.(2007)]{Tarter07} Tarter, J.~C., Backus, 
P.~R., Mancinelli, R.~L., et al.\ 2007, Astrobiology, 7, 30 
	\ifpp{\doi{10.1089/ast.2006.0124}}

\bibitem[Tian(2014)]{tian2014} Tian, F.  {\it Stability and Oxygen 
Contents of the Atmospheres of Planets in the HZ of M dwarfs},
Search for Life Beyond the Solar System: Exoplanets,
Biosignatures \& Instruments
	\ifpp{\ads{2014ebi..conf..2.5T}}

\bibitem[Tusnski \& Valio(2011)]{tusnski11} Tusnski, L.~R.~M., \&
Valio, A.\ 2011, \apj, 743, 97
	\ifpp{\doi{10.1088/0004-637X/743/1/97}}

\bibitem[Van Cleve \& Caldwell(2009)]{KIH09} Van Cleve, J.E., \& Caldwell, D.A.\ 2009,
{\it Kepler Instrument Handbook} (KSCI-19033)
	\ifpp{\url{https://archive.stsci.edu/kepler/documents.html}}

\bibitem[Von Korff et al.(2013)]{vonkorff13} Von Korff, J., 
Demorest, P., Heien, E., et al.\ 2013, \apj, 767, 40 
	\ifpp{\doi{10.1088/0004-637X/767/1/40}}

\bibitem[Weiss \& Marcy(2014)]{weiss_marcy14} Weiss, L.~M., \& 
Marcy, G.~W.\ 2014, \apjl, 783, L6 
	\ifpp{\doi{10.1088/2041-8205/783/1/L6}}

\bibitem[Wright(1992)]{wright92}Wright, J.~L. 1992, Space Sailing
(Philadelphia, PA: Gordon and Breach Science Publishers)
	\ifpp{\ads{1992spsa.book.....W}}

\bibitem[Wright et al.(2011)]{wright2011} Wright, J.~T., Fakhouri, 
O., Marcy, G.~W., et al.\ 2011, \pasp, 123, 412 
	\ifpp{\doi{10.1086/659427}}

\bibitem[Wright et al.(2014a)]{wright14a} Wright, J.~T., Mullan, 
B., Sigurdsson, S., \& Povich, M.~S.\ 2014, \apj, 792, 26
\ifpp{\doi{10.1088/0004-637X/792/1/26}}

\bibitem[Wright et al.(2014b)]{wright14b} Wright, J.~T., Griffith, 
R.~L., Sigurdsson, S., Povich, M.~S., \& Mullan, B.\ 2014, \apj, 792, 27 
\ifpp{\doi{10.1088/0004-637X/792/1/27}}


\bibitem[Zombeck(1990)]{zombeck} Zombeck, M.~V.\ 1990, Handbook of Space 
Astronomy and Astrophysics (2nd ed.; Cambridge: University Press)
	\ifpp{\ads{1990hsaa.book.....Z}}

\bibitem[Zsom et al.(2013)]{zsom2013} Zsom, A., Seager, S., de 
Wit, J., \& Stamenkovi{\'c}, V.\ 2013, \apj, 778, 109 
	\ifpp{\doi{10.1088/0004-637X/778/2/109}}

\bibitem[Zuluaga et al.(2013)]{zuluaga2013}Zuluaga, J.~I., 
Bustamante, S., Cuartas, P.~A., \& Hoyos, J.~H.\ 2013, \apj, 770, 23 
	\ifpp{\doi{10.1088/0004-637X/770/1/23}}

\end{thebibliography}
\end{document}